\documentclass[aps,a4,twocolumn]{revtex4-1}
\usepackage{amsmath}
\usepackage{latexsym}
\usepackage{float}
\usepackage{amssymb}
\usepackage{graphicx}
\usepackage{epsfig}

\begin{document}

\title{Conformational Studies of bottle-brush polymers absorbed on a flat solid surface}

\author{Hsiao-Ping Hsu$^a$, Wolfgang Paul$^b$, and Kurt Binder$^a$}
\affiliation{$^a$Institut f\"ur Physik, Johannes Gutenberg-Universit\"at Mainz,\\
 Staudinger Weg 7, D-55099 Mainz, Germany \\
$^b$Theoretische Physik, Martin Luther Universit\"at \\
Halle-Wittenberg, von Seckendorffplatz 1, 06120 Halle, Germany}

%\date{\today}
\begin{abstract}
The adsorption of a bottle-brush polymer end-grafted with one
chain end of its backbone to a flat substrate surface is studied
by Monte Carlo simulation of a coarse-grained model, that
previously has been characterized in the bulk, assuming a dilute
solution under good solvent conditions. Applying the bond
fluctuation model on the simple cubic lattice, we vary the
backbone chain length $N_b$ from $N_b=67$ to $N_b = 259$ effective
monomeric units, the side chain length $N$ from $N=6$ to
$N=48$, and the grafting density $\sigma=1$, i.e., parameters that correspond
well to the experimentally 
accessible range. When the adsorption energy strength $\epsilon$
is varied, we find that the adsorption transition (which becomes
well-defined in the limit $N_b \rightarrow \infty$, for arbitrary
finite $N$) roughly occurs at the same value $\epsilon_c$ as for
ordinary linear chains ($N=0$), at
least within our statistical errors. Mean square end-to-end
distances and gyration radii of the side chains are obtained, as
well as the monomer density profile in the direction perpendicular
to the adsorbing surface. We show that for longer side chains the
adsorption of bottle-brushes is a two step process, the decrease
of the perpendicular linear dimension of side chains with
adsorption energy strength can even be non-monotonic. Also the
behavior of the static structure factor $S(q)$ is analyzed,
evidence for a quasi-two-dimensional scaling is presented, and
consequences for the interpretation of experiments are discussed.
\end{abstract}

\maketitle

\section{Introduction}
Macromolecules with a comb-like chemical architecture, where
flexible side chains are densely grafted to a (also flexible)
polymer acting as ``backbone'' have found abiding current
interest, see, e.g., the recent reviews \cite{1,2,3,4}. These
so-called ``bottle-brush polymers'' exhibit an interesting
competition due to the steric repulsion between the side chains
and the configurational entropy of the backbone: varying the
grafting density of the side chains and their length, the
effective stiffness of these cylindrical brushes can be controlled
over a wide range. Since the overall conformation of the
bottle-brush polymer is sensitive to external stimuli such as
solvent quality, pH value and ionic strength of the solution, or
temperature, electromagnetic fields, various possible
applications of these bottle-brush polymers have been discussed
(actuators, sensors, building blocks of new nanostructures or
templates for producing metallic nano-wires.
\cite{5,6,7,8,9}). A fascinating aspect is also the importance of
biomolecules with bottle-brush architecture, such as proteoglycans
\cite{10} (polyelectrolytes consisting of a protein backbone and
carbohydrate side chains, performing biological functions from
cell signaling and cell surface protection to joint lubrication
\cite{11,12}). Of course, under many circumstances these functions
of bottle-brushes occur when they are attached to a substrate
surface (e.g. a cell membrane in a biological context, or an
inorganic flat solid surface for nano-technological applications,
including also special surface coatings \cite{13}). Bottle-brush
molecules attached to surfaces are also of particular interest,
since additional experimental tools become available to study
their structure (e.g., one can directly visualize their
large-scale conformation by scanning force microscopy
\cite{3,14,15}, or one can use atomic force microscopes to measure
force vs. extension curves \cite{16}). Thus, while rich
information on the structure of bottle-brush polymers at surfaces
already is available \cite{3}, it is not so clear which of these
features are intrinsic properties of these complex macromolecules,
and which features are only induced by their adsorption to the
substrate. We recall that basic aspects of bottle-brush polymers
in dilute solution away from any constraining surfaces are still
incompletely understood (e.g. the relation between the persistence
length of these macromolecules and their microscopic
characteristics such as side-chain and backbone chain lengths $N$
and $N_b$ has been controversial, see e.g. \cite{17,18,19}). There
are already some indications from theoretical work \cite{20,21}
that strongly adsorbed bottle-brushes exhibit properties different
from the bulk.

Thus the present paper intends to contribute to a clarification of
this problem, by presenting a comprehensive computer simulation
study of (large) bottle-brush polymers near surfaces, over the
full range extending from a repulsive surface (applying only the
constraint that one backbone chain end is grafted to this surface)
through the region of the adsorption transition, where the overall
conformation of the macromolecule changes from a three-dimensional
``mushroom'' to a quasi-two-dimensional ``pancake'', to the
strongly adsorbed almost two-dimensional case. Using side chain
lengths up to $N=48$ we ensure that statistical concepts of
theoretical polymer physics \cite{22} start to become applicable
\cite{23}, and we also note that our range of side chain lengths
can be nicely mapped to experiments \cite{19}. While the
adsorption transition of flexible linear chains has been studied
extensively by various theoretical methods \cite{24,25,26,27,28,29,30,31,31a},
the present work is the first one 
to present large-scale simulations on the adsorption of
bottle-brush polymers. Clearly, the interplay of the
conformational changes due to adsorption with the chain stiffness
(tunable via the side chain length variation) is of particular
interest \cite{32}.

The plan of this paper is as follows: Section II describes the
model and the simulation techniques while Section III presents the
numerical results. Section IV gives a summary and an outlook on
both pertinent experimental work and on open questions.

\section{MODEL AND SIMULATION TECHNIQUE}
The bond fluctuation model on the simple cubic lattice
\cite{33,34,35,36} has been extended in previous work to simulate
bottle-brush polymers in dilute solution under good solvent
conditions \cite{19,37}. In the present paper, we extend these
studies to bottle-brush polymers tethered with one chain end of
their backbone to a flat impenetrable surface, which is
chosen to be the xy-plane (z = 0) of the lattice. Besides the
excluded volume interactions between the monomers of the side
chains and/or the backbone chain, and the infinitely repulsive
interaction between monomers and the grafting surfaces (which
prevents that monomers can occur with z-coordinates $z <0$ in the
system), we also consider the effect of an additional attractive
short range interaction $\epsilon$ between the grafting surface
and the monomers of the bottle-brush polymer. Recall that in the
bond fluctuation model each effective monomer blocks all 8 sites
at the corners of an elementary cube of the lattice from further
occupation. Thus, an energy $\epsilon$ is won if 4 sites of such a
cube are in plane $z=0$, while the remaining 4 sites of the cube
then have to be in the adjacent plane $z=1$ (note that we use the
lattice spacing as the unit of length).

\begin{figure*}[htb]
\begin{center}
\includegraphics[scale=0.29,angle=0]{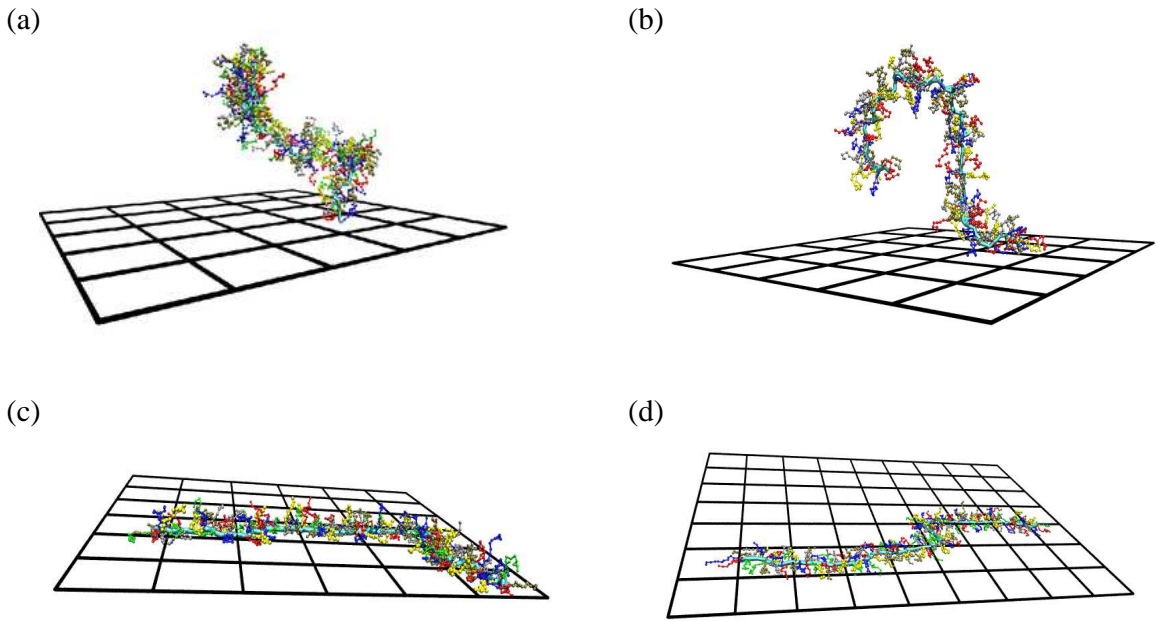}
\caption{\label{fig1} Snapshots of typical
configurations of bottle-brush polymers for backbone chain length
$N_b=131$ and side chain length $N=12$, and four choices of
adsorption energy $\epsilon$, namely $\epsilon =0$ (a), 1.0 (b),
1.5 (c) and 2.0 (d).}
\end{center}
\end{figure*}

\begin{figure*}
\begin{center}
\includegraphics[scale=0.30, angle=270]{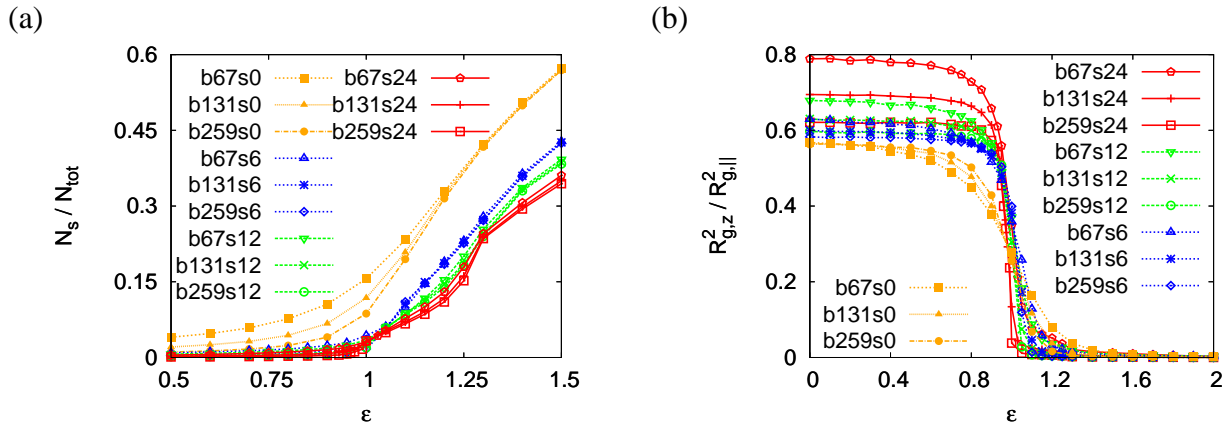}
\caption{\label{fig2} (a) Fraction of
monomer-surface contracts $N_s/N_{\rm tot}$ and (b) ratio
$R^2_{g,z}/R^2_{g,||}$ of the mean square gyration radii of the
whole bottle-brush polymers perpendicular and parallel to the
surface plotted against the adsorption energy $\epsilon$. Data for
backbone lengths $N_b = 67, 131$ and 259 are included, as well as
side chain length $N=0, 6, 12$ and 24.}
\end{center}
\end{figure*}

The bond vectors connecting two adjacent monomers along a chain
are chosen from the set \{[2,0,0], [2,1,0], [2,1,1], [2,2,1], [3,0,0],
and [3,1,0]\}, including also all possible rotations, reflections
and reversions of these bond vectors. We use the same set of bond
vectors, irrespective of whether we consider side chains of the
bottle-brush or the backbone chain. Note that this model is one of
the standard models used for Monte Carlo simulation of
macromolecular systems, many physical properties of polymers are
rather well accounted for, and very efficient simulation
algorithms can be formulated for this model \cite{38}.

The chemical architecture of a bottle-brush polymer is arranged
such that at a backbone monomer that acts as a grafting site one
side chain is anchored, and the chemical distance (i.e.,
difference in the consecutive labels of the backbone monomers) is
the inverse of the grafting density, $1/\sigma$. Thus, only
integer numbers $1/\sigma$ are possible, and side-chains are
anchored at regular chemical distances. Note that both next to the
first grafting site and next to the last grafting site of the
backbone we add one extra backbone monomer. Thus if we have $n_c$
side chains, the total number of monomers in the backbone is
$N_b=[(n_c-1)/\sigma + 1]+2$, and the total number of monomers in
the bottle-brush then is $N_{\textrm{tot}}=N_b +n_cN$, when $N$
denotes the number of monomers per side chain. Now, we also
require that one of the two extra monomers at a chain end of the
backbone is attached to the adsorbing surface at $z=0$, and hence
our simulation deals with a ``bottle-brush mushroom'' (recall that
an isolated polymer chain \cite{24}, end-grafted to a planar
surface traditionally is called a ``polymer mushroom'', and we
extend this nomenclature to bottle-brushes). In the present
work, only $\sigma=1$ is considered.

The initial configuration of the bottle-brush polymer is
constructed by assuming that the structures of the backbone chain
and of the side chains are both rod-like. Namely, the backbone is
oriented in the direction along the + z-direction, setting all
bond vectors between monomers equal to $\vec{r}= 3 \hat{z}$,
where $\hat{z}$ is a unit vector along the z-axis. The bond
vectors between monomers on each side chain are chosen randomly
from one of the allowed bond vectors in the x-y plane. This
initial configuration, of course, is very far from equilibrium,
and needs to be carefully relaxed towards equilibrium in the first
part of the Monte Carlo run.

In order to speed up the dynamics of the model, we do not move the
monomers from their previous positions to only one of the six
nearest neighbor sites of the monomer, as is traditionally done,
but allow attempts to move a monomer to one of the 26 sites
surrounding the current position of the monomer. This ``L26 move''
allows bonds to cross each other and has been shown to yield a
significant speedup of the algorithm in comparison with the
standard ``L6'' move \cite{38}. In addition to the local moves,
the pivot algorithm \cite{39} is employed in the following ways:
(a) A monomer on the backbone is chosen randomly and the part of
the bottle-brush polymer containing the free end of the backbone is
exposed to a trial move, by randomly choosing one operation from
the 48 symmetry operations of the lattice (no change, rotations by
$ 90^\circ$ and $180^\circ$ angles, reflections, and inversions).
Of course, the trial configuration is accepted as a new
configuration only, if it does not violate any excluded volume
constraints and if it passes the Metropolis acceptance test (where
the change in adsorption energy enters). The same is true for the
attempted ``L26'' moves, of course. (b) A monomer is chosen
randomly from the monomers on all side chains, and the trial move
is effected by transforming the part of the side chain from the
selected monomer to the free end by one of the
same 48 symmetry operations as mentioned above.

A Monte Carlo step (MCS) in our simulation then consists of a
sequence of $N_{\textrm{tot}}$ ``L26'' moves (each monomer on
average is attempted to be moved once), $k_{pb}$ pivot moves of
choosing a backbone monomer randomly, and $k_{ps}$ pivot moves of
choosing a side chain monomer randomly. We set $k_{ps}=n_c/4$ and
adjusted $k_{pb}$ such that the acceptance ratio was about 30 \%
to 40\% for small values of $\epsilon$. However, when a
bottle-brush polymer becomes strongly adsorbed to the surface
successful pivot moves are more difficult. To avoid the problem
that in the regime of the strong adsorption the bottle-brush
configurations are frozen, 16 to 512 different equilibrium
configurations are generated as initial configurations for the
start of Monte Carlo simulations in equilibrium. During the
equilibration process, the convergence of the time series for the
energy and the gyration radius components of the bottle-brush
polymer are monitored. For the measurements at each parameter set
in total about 10$^5$ to 10$^6$ statistically independent
configurations were generated. Fig.~\ref{fig1} shows typical
snapshot pictures of such configurations for 4 choices of
$\epsilon$.

\begin{figure*}[htb]
\begin{center}
\includegraphics[scale=0.30,angle=270]{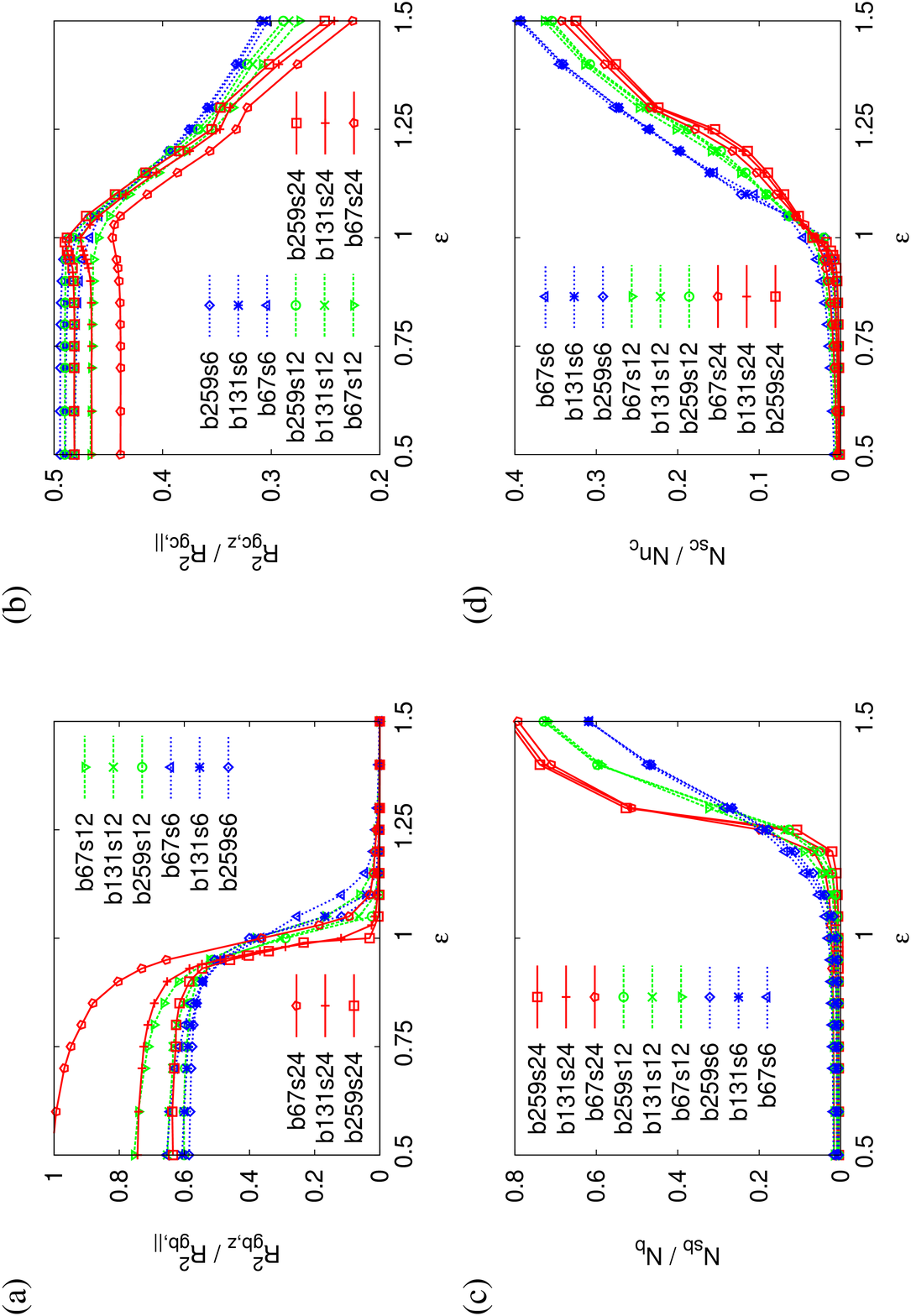}
\caption{\label{fig3} Ratio between the mean square
gyration radius component perpendicular and parallel to the
surface plotted vs. the adsorption energy $\epsilon$ for the
bottle-brush backbone, $R_{gb,z}^2/R_{gb,||}^2$ (a) and for the
side chains, $R^2_{gc,z}/R^2_{gc,||}$(b). Fraction of monomer
surface contacts plotted vs. adsorption energy $\epsilon$ for the
backbone of the bottle-brush, $N_{sb}/N_b$ (c) and for the side
chains, $N_{sc}/(Nn_c)$; (d).}
\end{center}
\end{figure*}

\begin{figure*}[htb]
\begin{center}
\includegraphics[scale=0.29,angle=270]{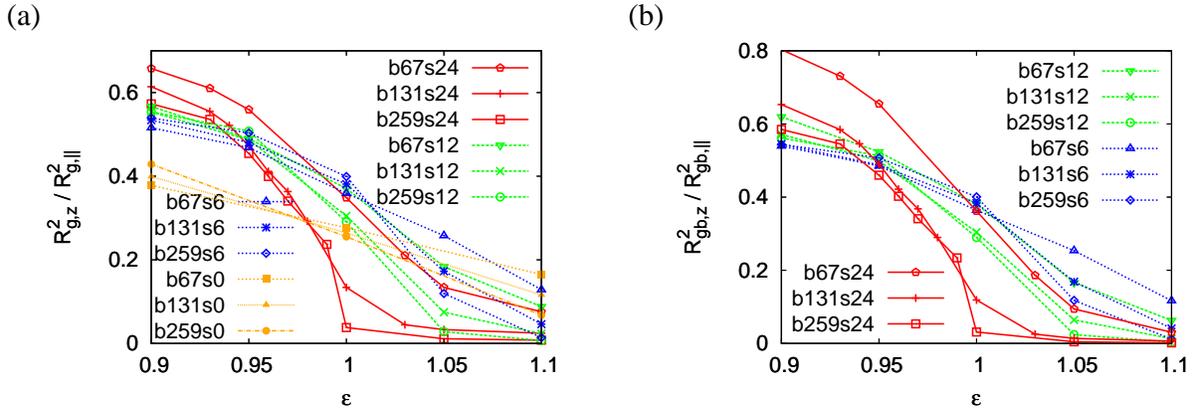}
\caption{\label{fig034} Same figures as shown in
Fig.~\ref{fig2}(b) and Fig.~\ref{fig3}(a) but for the data
in the range $0.9 < \epsilon <1.1$.
For $N=12$ and $N=24$ the backbone chain length $N_b=67$ is too
short so the expected scaling behavior (which requires an
intersection of the curves at $\epsilon_c$) is not yet seen.}
\end{center}
\end{figure*}

\begin{figure*}[htb]
\begin{center}
\includegraphics[scale=0.30,angle=270]{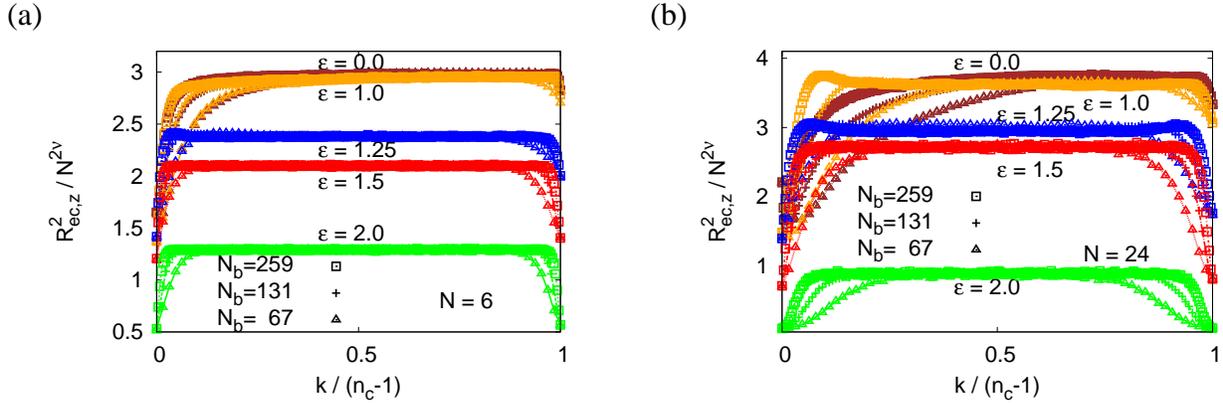}
\caption{\label{fig4} Rescaled mean square
end-to-end distance of the side chains, perpendicular to the
surface, $R_{ec,z}^2/N^{2 \nu}$, vs. the normalized position
$k/(n_c-1)$ in the coordinate system along the backbone (the $n_c$
side chains are labeled $k=0,1,\ldots,n_c-1$, the side chain with
$k=0$ is next to the grafted backbone monomer, while the side
chain with $k=n_c-1$ is next to the backbone free end). Five
choices of adsorption energy $\epsilon =0$, $1.0$, $1.25$, $1.5$
and $2.0$
are included, as indicated, as well as three choices of the
backbone chain length, $N_b=67$, $131$ and $259$, respectively. Case
(a) refers to $N=6$ and case (b) to $N=24$.}
\end{center}
\end{figure*}

\begin{figure*}[htb]
\includegraphics[scale=0.30,angle=270]{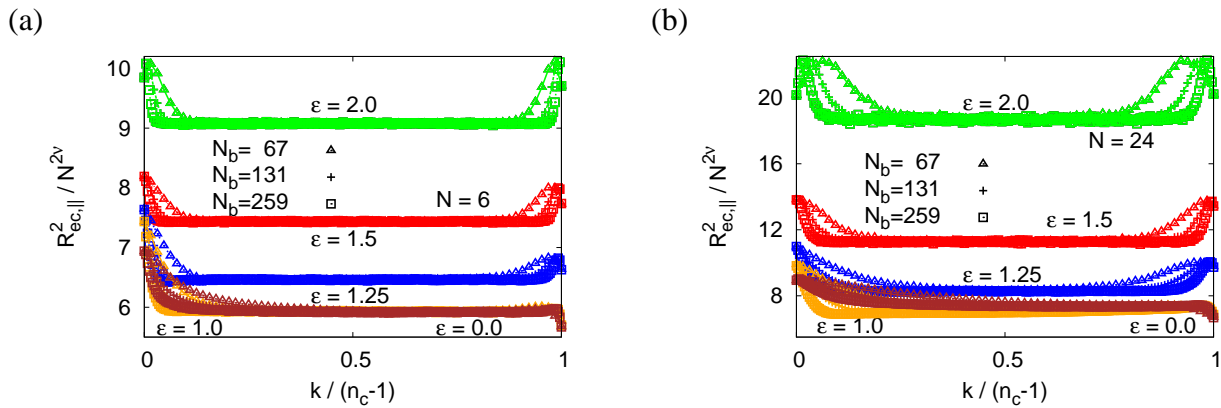}
\begin{center}\caption{\label{fig5} Same as Fig.~\ref{fig4}, but
for the component parallel to the surface, $R_{ec,||}^2/N^{2
\nu}$, plotted vs. $k/(n_c-1)$.}
\end{center}
\end{figure*}

\begin{figure*}[htb]
\begin{center}
\includegraphics[scale=0.30,angle=270]{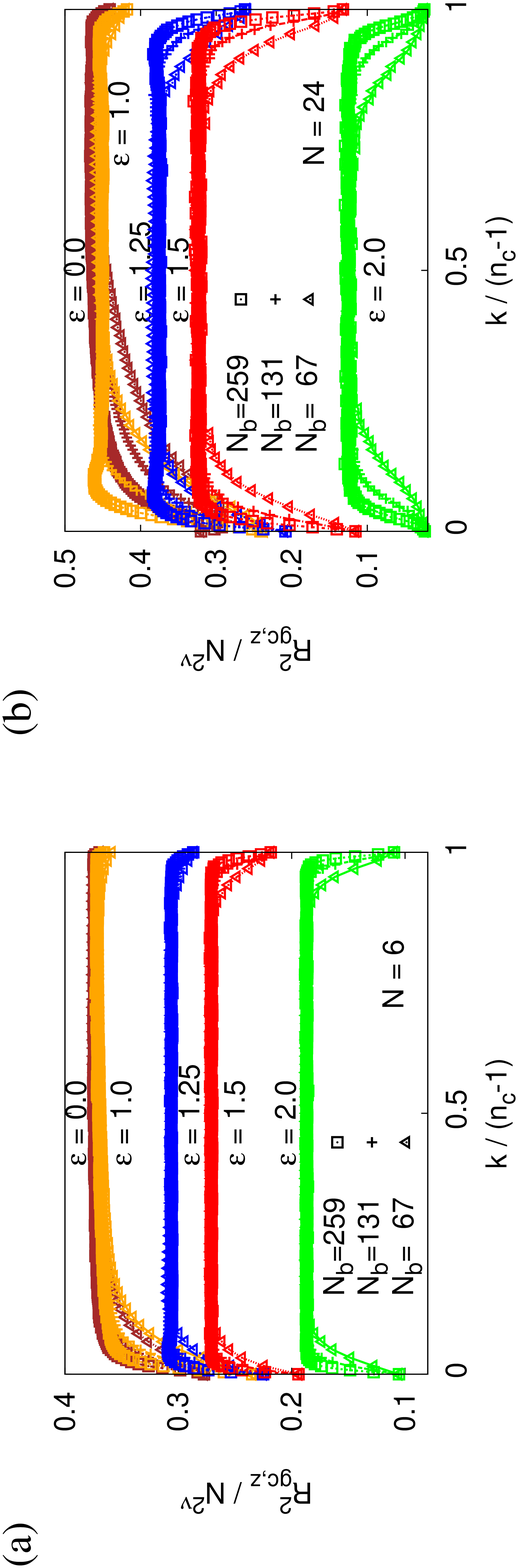}
\caption{\label{fig6} Same as Fig.~\ref{fig4}, but
for the rescaled mean square gyration radius of the side chains,
perpendicular to the surface, $R_{gc,z}^2/N^{2 \nu}$, for $N=6$
(a) and $N=24$ (b).}
\end{center}
\end{figure*}

\section{RESULTS}
When a single flexible polymer chain of chain length $N_b$
tethered with one end to a flat impenetrable surface experiences
an energy $\epsilon$ that monomers win when they are adjacent to
the surface, a transition of this ``polymer mushroom'' occurs from
an essentially three-dimensional configuration to a
quasi-two-dimensional ``pancake'' configuration when the
adsorption energy exceeds a critical value $\epsilon_c$
\cite{24,25,26,27,28,29,30,31,31a}. This transition is a second order
transition, involving a nontrivial exponent $\phi$ (the
``crossover exponent'' \cite{26}) such that the fraction of
monomer-surface contacts $N_s/N_b$ behaves as (in the limit
$N_b \rightarrow \infty$) \cite{26}

\begin{eqnarray}\label{eq1}
\frac{N_s}{N_b} \propto \left\{
\begin{array}{ll}
\frac{1}{N_b}(1-\epsilon/\epsilon_c)^{-1}, & \epsilon < \epsilon_c \\
N_b^{\phi-1}, & \epsilon = \epsilon_c  \\
(\epsilon/\epsilon_c - 1)^{1/\phi-1},  &\epsilon > \epsilon_c
\end{array} \right. 
\end{eqnarray}
Despite a lot of effort (see \cite{28,29,30,31,31a} and references
therein) the value of the (universal) exponent $\phi$ is not yet
very precisely known, some estimates being compatible with
$\phi = 0.50 \pm 0.02$ \cite{29,30}. 
However, the most recent estimate from Monte Carlo simulations
of the bond fluctuation model yielded~\cite{31a} $\phi=0.59$,
a value close to the early estimate~\cite{26} $\phi=0.58$.
This adsorption transition also
shows up in the chain linear dimensions, of course. While for
$\epsilon \leq \epsilon_c$ all linear dimensions are of the same
order \cite{26,28}
\begin{equation}\label{eq2}
R_{g,z}^2\propto R_{g,||}^2 \propto N_b^{2\nu}\quad , \;\nu
\approx 0.588\quad ,
\end{equation}
for $\epsilon > \epsilon _c$ the perpendicular component of the
gyration radius (related to the thickness of the ``pancake''
configuration) remains finite \cite{26},
\begin{equation}\label{eq3}
R_{g,z}^2 \propto (\epsilon /\epsilon_c - 1) ^{-2 \nu/\phi} \; ,
\end{equation}
while the parallel component exhibits a scaling compatible with
two-dimensional self-avoiding walks \cite{26},
\begin{equation}\label{eq4}
R_{g,||}^2 \propto (\epsilon/\epsilon_c-1)^{2 (\nu_2-\nu)/\phi}
N_b^{2\nu_2} \; , \nu_2 = 3/4\;.
\end{equation}
Hence when one plots the ratio $R^2_{g,z}/R^2_{g,||}$ vs.
$\epsilon$ for several finite large values of $N_b$, one should
see a family of curves which exhibit an intersection point for
$\epsilon = \epsilon_c$.

Motivated by Eqs.~(\ref{eq1}), (\ref{eq3}) and (\ref{eq4}), we plot the
fraction of monomer-surface contacts $N_s/N_{\textrm{tot}}$ versus
$\epsilon$ (Fig.~\ref{fig2}(a)), $N_{\textrm{tot}}$ being the total
number of effective monomers in the bottle-brush polymer, as well
as the ratio $R^2_{g,z}/R^2_{g,||}$ (Fig.~\ref{fig2}(b)). Inspection
of these ``raw data'' (Figs.~\ref{fig2},\ref{fig3})
already indicates that the adsorption
transition occurs near $\epsilon_c(N) \approx 1.00 \pm 0.05$
(see Fig.~\ref{fig034}, a close view of the ratio $R^2_{g,z}/R^2_{g,||}$
in the vicinity of $\epsilon= 1$); 
if $\epsilon_c(N)$ depends at all on side chain length $N$, the
dependence is rather weak. However, it is remarkable that
$N_s/N_{\textrm{tot}}$ increases for $\epsilon > \epsilon_c(N)$
clearly much faster for the linear chains $(N=0)$ than for
the bottle-brush polymers; $N_s/N_{\textrm{tot}}$ seems to
converge to a limiting function for large $N$ that is independent
of $N$. In contrast, the ratio $R_{g,z}^2/R^2_{g,||}$ varies more
steeply the larger $N$, and there is no universal behavior; in
particular, in the non-adsorbing regime $R^2_{g,z}/R^2_{g,||}$
increases monotonically with $N$. This behavior is plausible,
since the longer the side chains are the more the excluded volume
interaction with the planar substrate surface is felt.

\begin{figure*}[htb]
\begin{center}
\includegraphics[scale=0.30,angle=270]{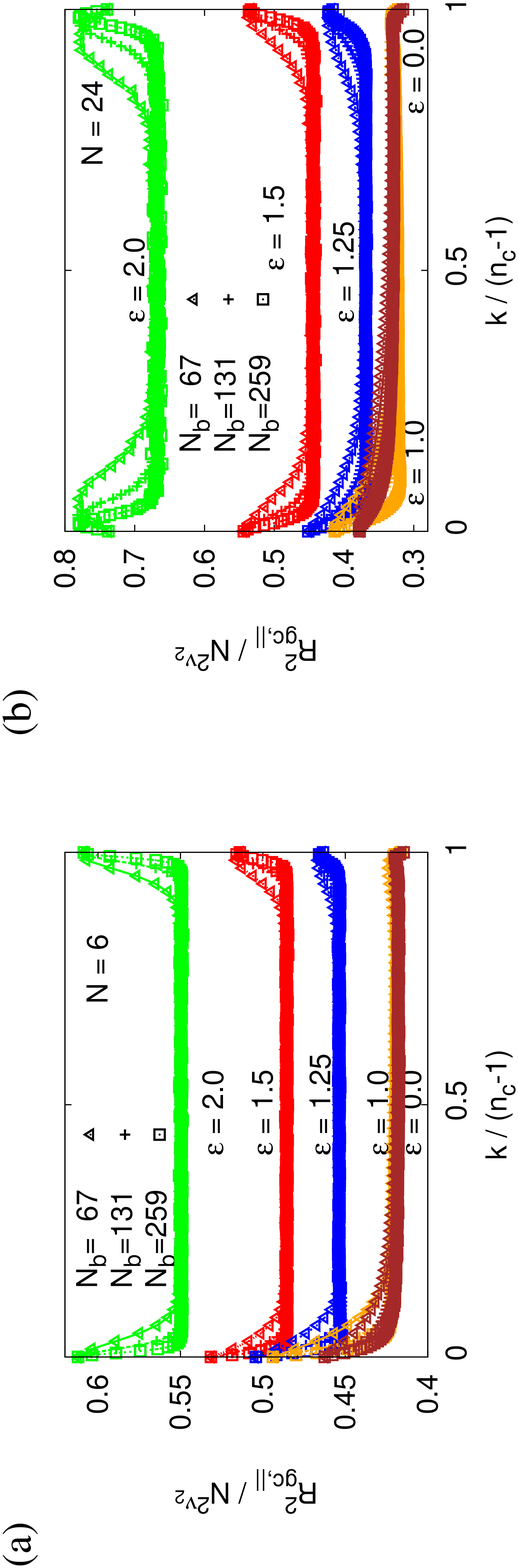}
\caption{\label{fig7} Same as Fig.~\ref{fig4}, but
for $R^2_{gc,||}/N^{2\nu_2}$, plotted vs. $k/(n_c-1)$. Note that
by the normalization with $N^{2\nu_2}$ rather than $N^{2 \nu}$ the
shown ratios approach a finite limit for $N \rightarrow \infty$,
unlike Fig.~\ref{fig5}.}
\end{center}
\end{figure*}

\begin{figure*}[htb]
\begin{center}
\includegraphics[scale=0.30,angle=270]{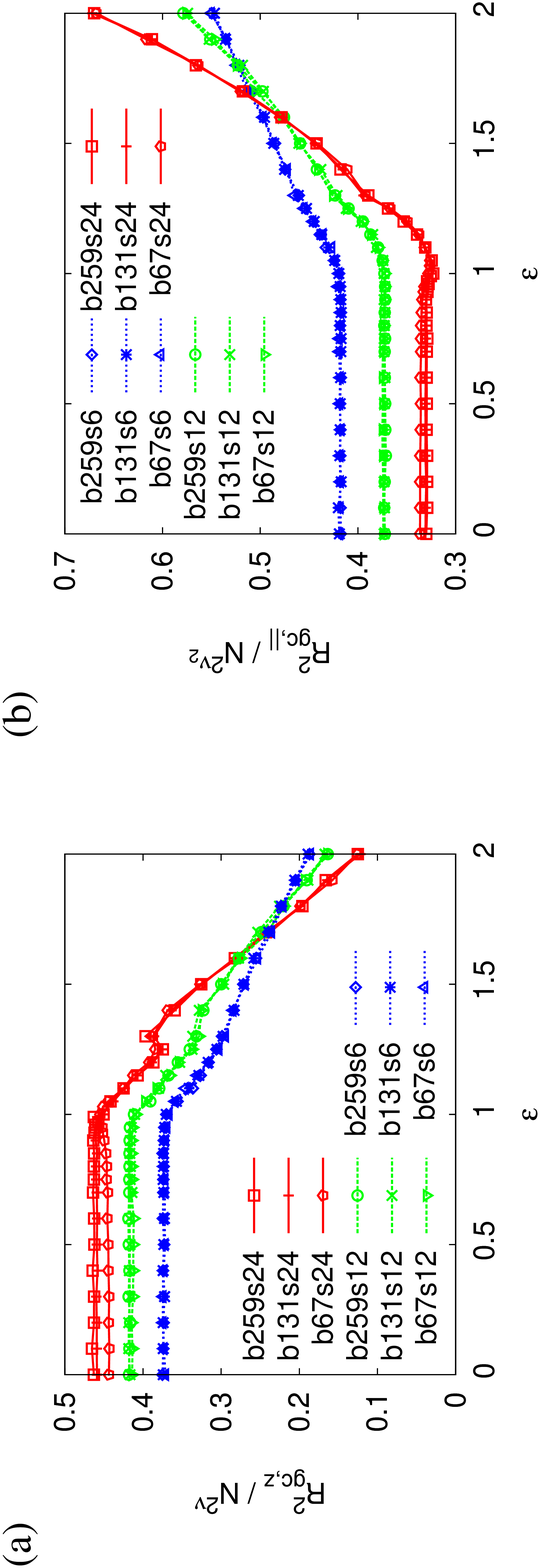}
\caption{\label{fig8} Rescaled mean square gyration
 radius $R^2_{gc,z}/N^{2\nu}$ (a) and $R^2_{gc,||}/N^{2 \nu_2}$ (b)
 plotted vs. $\epsilon$, including data for three different
 backbone chain lengths $N_b=67$, 131 and 259 as well as three
 different side chain lengths N = 6, 12 and 24 (the various
 combinations of $N_b$ and $N$ are indicated in the figure as
 $bN_bsN$, respectively). Note that these data, unlike
 Figs.~\ref{fig6}, \ref{fig7}, are averages along the backbone.}
\end{center}
\end{figure*}

Of course, it would be desirable to extract more precise estimates
of $\epsilon_c(N)$ from a scaling analysis of the data, similar
to the scaling analysis done in Refs.~\cite{26,29,31,31a}.
Unfortunately, as explained in the Appendix, no significant 
gain in accuracy in comparison with the simple intersection method
of Fig.~\ref{fig2}(b) could be gotten.
An obvious question with respect to the adsorption transition of
bottle-brush polymers is the clarification whether the behavior of
the side chains and of the backbone is fully analogous, or whether
some characteristic differences occur. Thus Fig.~\ref{fig3}
disentangles in both the fraction of adsorbed monomers and in the
ratio $R^2_{g,z}/R^2_{g,||}$ the contributions of the backbone from
the contributions of the side chains. One sees remarkable differences between
the behavior of the side chains and the backbone: while the ratio 
$R^2_{gc,z}/R^2_{gc,||}$ is essentially constant for $\epsilon \le \epsilon_c$
and then decreases only rather gradually (almost linearly), the ratio
$R^2_{gb,z}/R^2_{gb,||}$ shows most of its decrease before $\epsilon_c \approx
1.0$ is reached. On the other hand, for large $N$ a significant rise of
$N_{sb}/N_b$ only starts at about $\epsilon\approx 1.2 > \epsilon_c$.
Some aspects of this behavior can be attributed
to the fact that in the adsorption of bottle-brushes it easily
happens that the monomer density in the layer adjacent to the
adsorbing wall gets rather high, thus there is not enough empty
space in this layer to allow for the adsorption of all the
side-chain monomers.

Since some of the side chains 
near the backbone chain end which is grafted to the substrate
surface necessarily are rather close to the adsorbing surface,
while this is not true for the side chains in the vicinity of its free ends,
it also is interesting 
to resolve the side-chain properties as a function of their
position along the backbone (Fig.~\ref{fig4}). One can see that
the linear dimensions of the side chains near to the grafted 
backbone chain end are strongly reduced for non-adsorbed 
chains, while an
analogous effect near the free backbone end is much smaller. No
such asymmetry can occur for free bottle-brush polymers in bulk
solution, of course, since there both backbone chain ends are
strictly equivalent. It is remarkable, however, that this
asymmetry effect vanishes almost completely when the bottle-brush
polymer gets adsorbed: then also the free chain end of the
backbone is close to the surface, and it does not matter whether
it actually would be grafted or not. A similar asymmetry is found,
however, when one considers the mean square end-to-end distance of
the side chains parallel to the surface (Fig.~\ref{fig5}): of
course, then the magnitude of $R^2_{ec,||}/N^{2 \nu}$ strongly
increases with $\epsilon$, unlike $R_{ec,z}^2/N^{2 \nu}$, due to
the tendency of the adsorbed polymers to form ``pancake''
conformations, cf. Eqs.~(\ref{eq3}), (\ref{eq4}),
and therefore these ratios systematically increase with
increasing $N$.

Similar observations can be made when one studies the mean square
gyration radius components (Figs.~\ref{fig6}, \ref{fig7}). Thus,
if one would normalize the data for $R^2_{ec,||}$ with $N^{2
\nu_2}$ with $\nu_2 = 3/4$ instead of $N^{2 \nu}$ with $\nu =
0.588$, as done in Fig.~\ref{fig5}, one finds plateau values for
$\epsilon = 2.0$ in the center of the bottle-brush 
$(0.2 < k/(n_c-1)< 0.8)$ at about $5.1$ for N = 6 and at about 
$6.8$ for $N = 24$. These values are of the same order 
as seen in Fig.~\ref{fig5} for
$\epsilon \leq 1.0$, where the side chains still take the
configuration of three-dimensional coils. Of course, side chain
lengths $N \leq 48$, as studied here, are too short to reveal the
asymptotic behavior $R^2_{ec,||} \propto N^{2 \nu_2}$ in the
regime $\epsilon > \epsilon_c$ clearly.

\begin{figure*}[htb]
\begin{center}
\includegraphics[scale=0.30,angle=270]{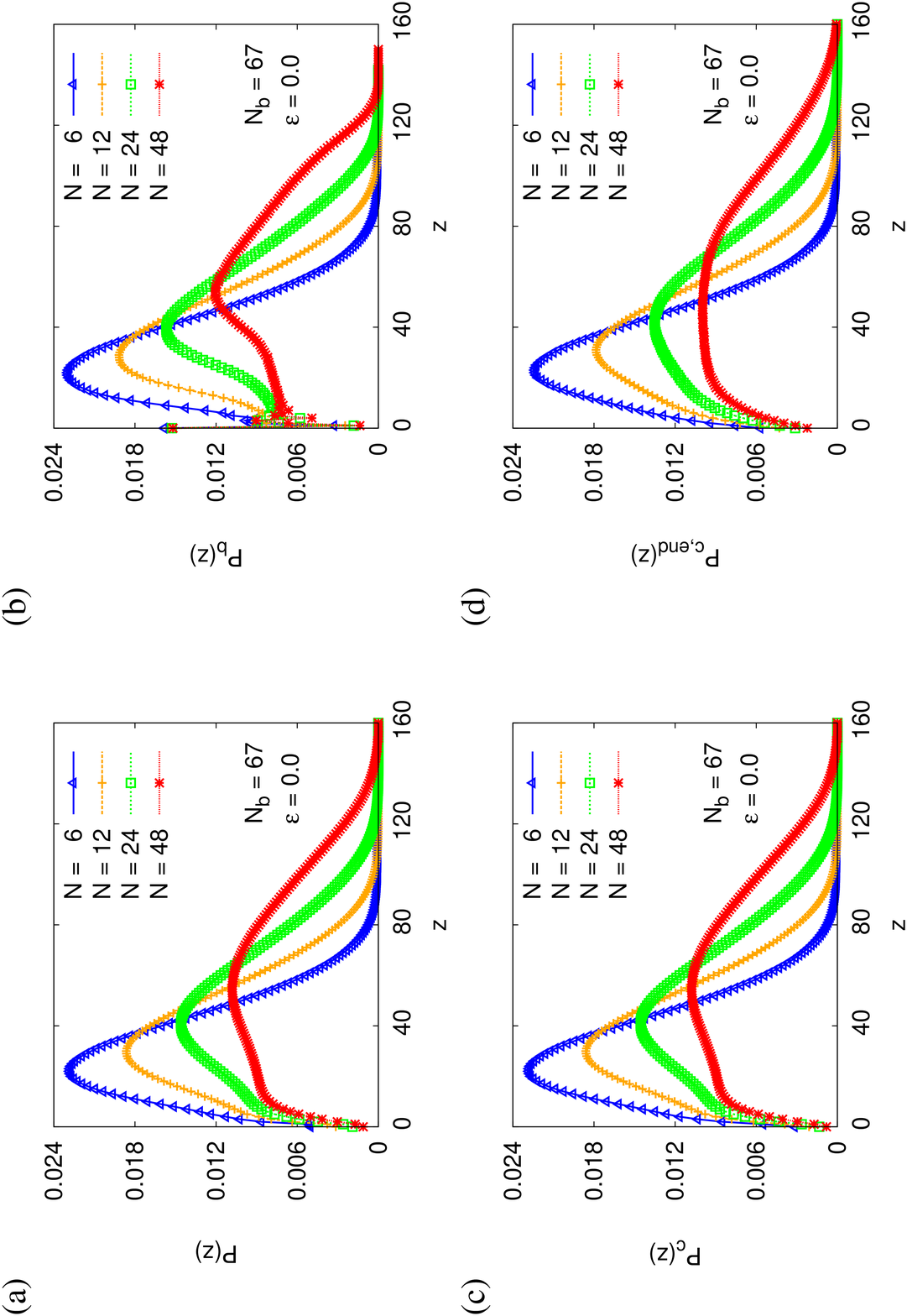}
\caption{\label{fig9} Monomer density profiles of
the bottle-brush polymers $P(z)$ as a function of the distance $z$
from the surface, for backbone chain length $N_b= 67$ at the
non-adsorbing case ($\epsilon = 0.0$) and four side chain lengths,
N = 6, 12, 24 and 48. Case (a) refers to all monomers of the
bottle-brush, case (b) includes monomers of the backbone only,
case (c) includes only the side chains, while case (d) includes
only the free ends of the side chains.}
\end{center}
\end{figure*}

\begin{figure*}[htb]
\begin{center}
\includegraphics[scale=0.30,angle=270]{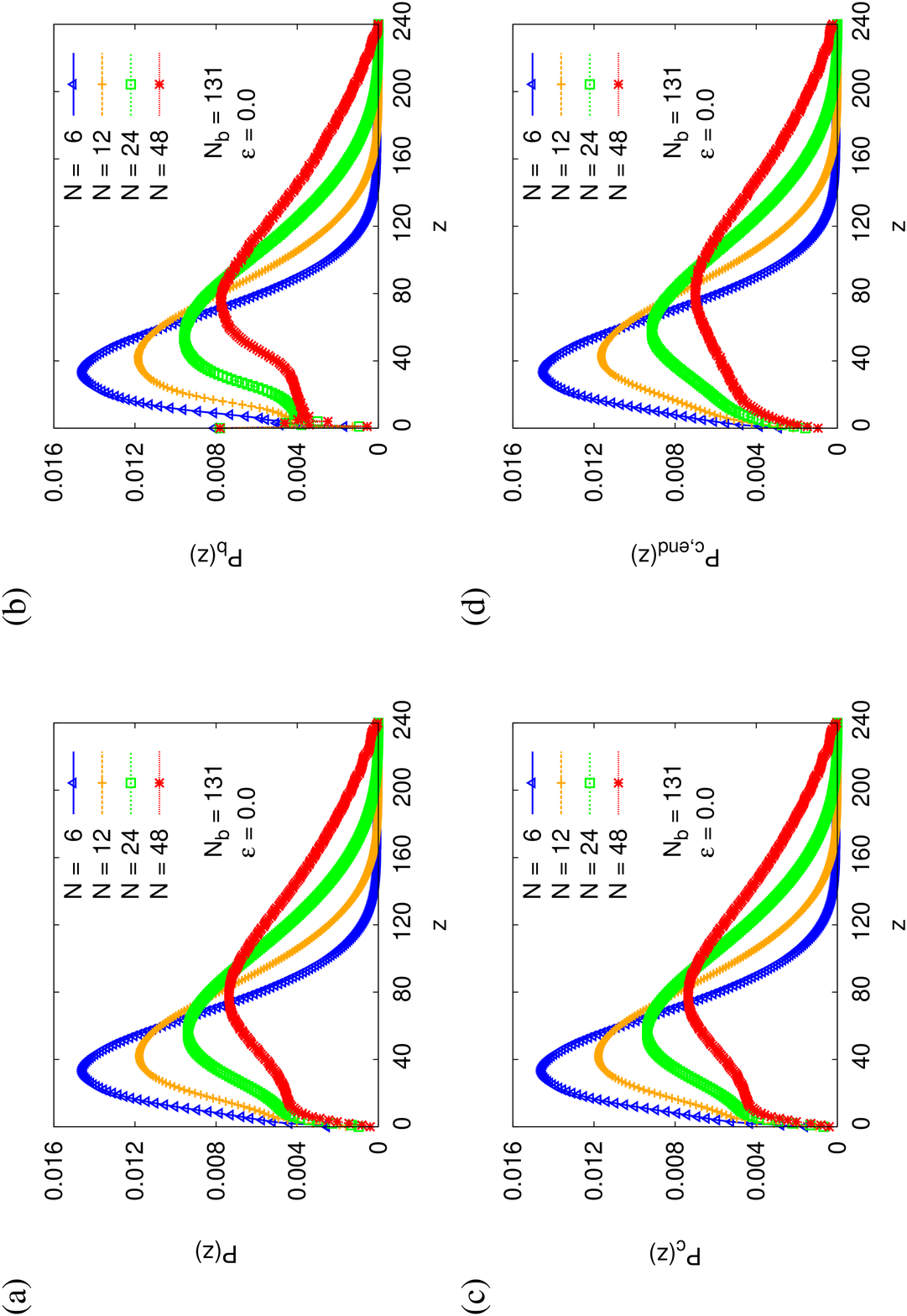} \hspace{0.4cm}
\caption{\label{fig10} Same as Fig.~\ref{fig9}, but
for $N_b=131$ and $\epsilon = 0.0$}
\end{center}
\end{figure*}

\begin{figure*}[htb]
\begin{center}
\includegraphics[scale=0.30,angle=270]{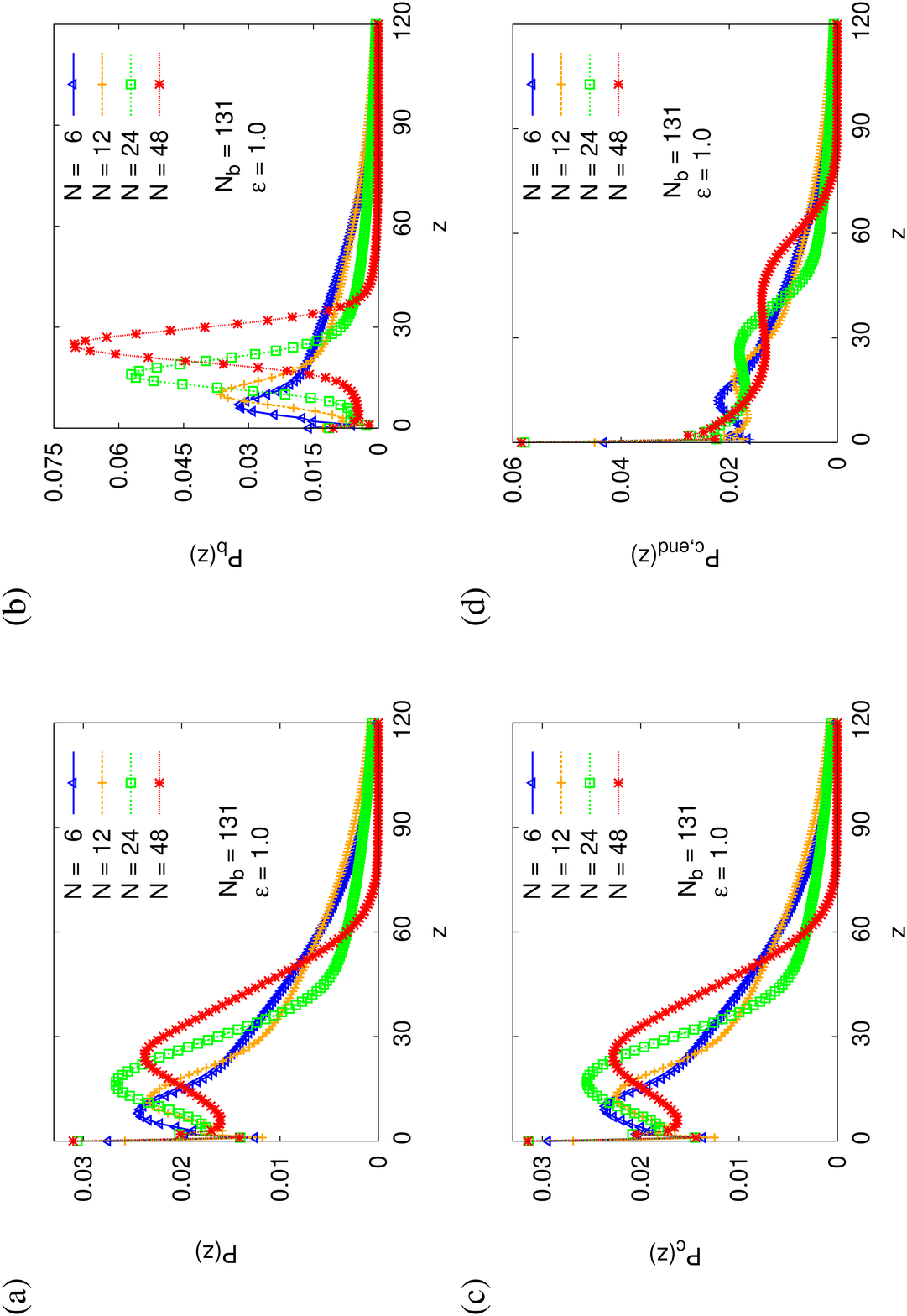} 
\caption{\label{fig11} Same as Fig.~\ref{fig9}, but
for $N_b= 131$ and $\epsilon = 1.0$}
\end{center}
\end{figure*}

\begin{figure*}[htb]
\begin{center}
\includegraphics[scale=0.30,angle=270]{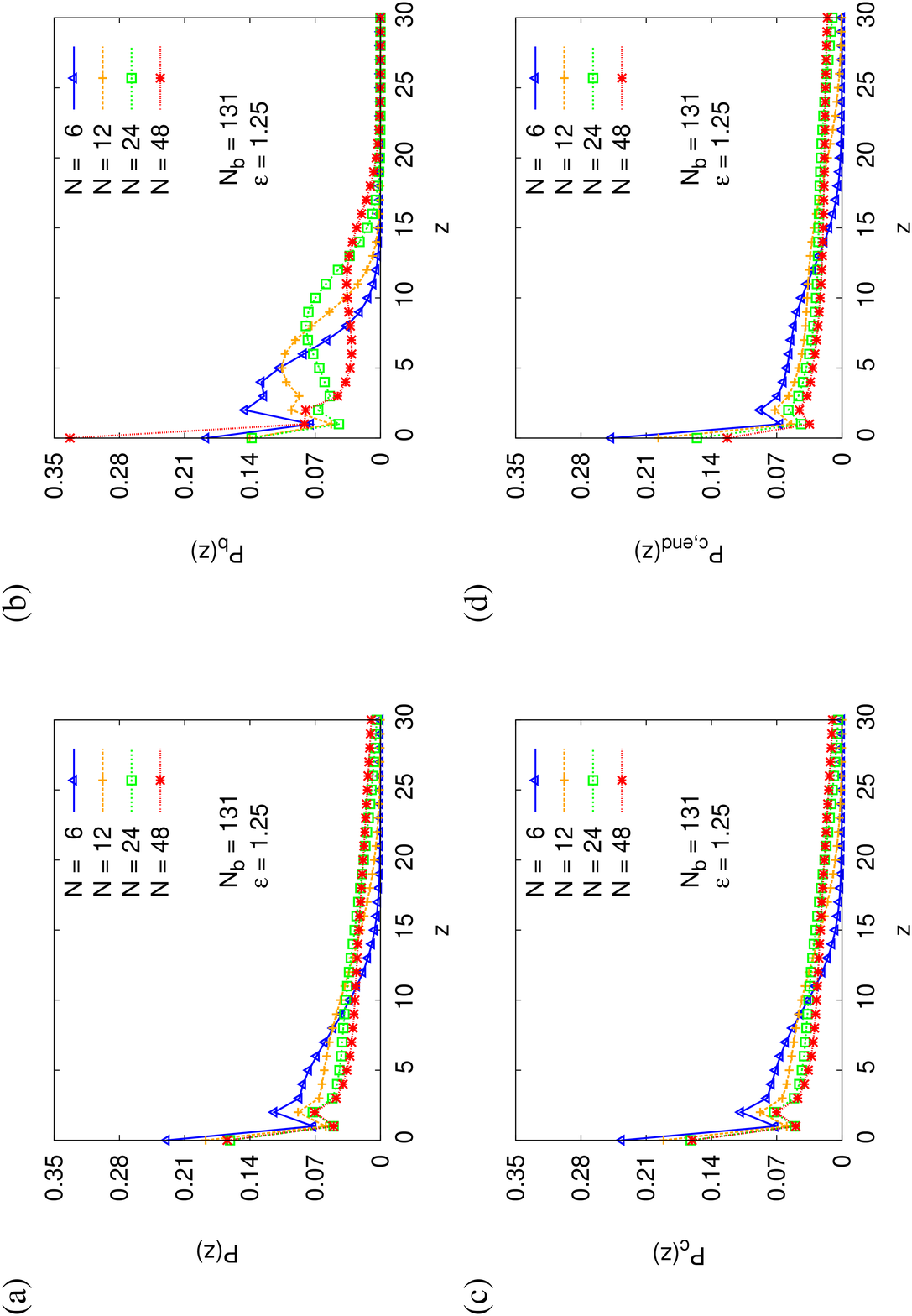}
\caption{\label{fig12} Same as Fig.~\ref{fig9}, but
for $N_b = 131$ and $\epsilon = 1.25$}
\end{center}
\end{figure*}

\begin{figure*}[htb]
\begin{center}
\includegraphics[scale=0.30,angle=270]{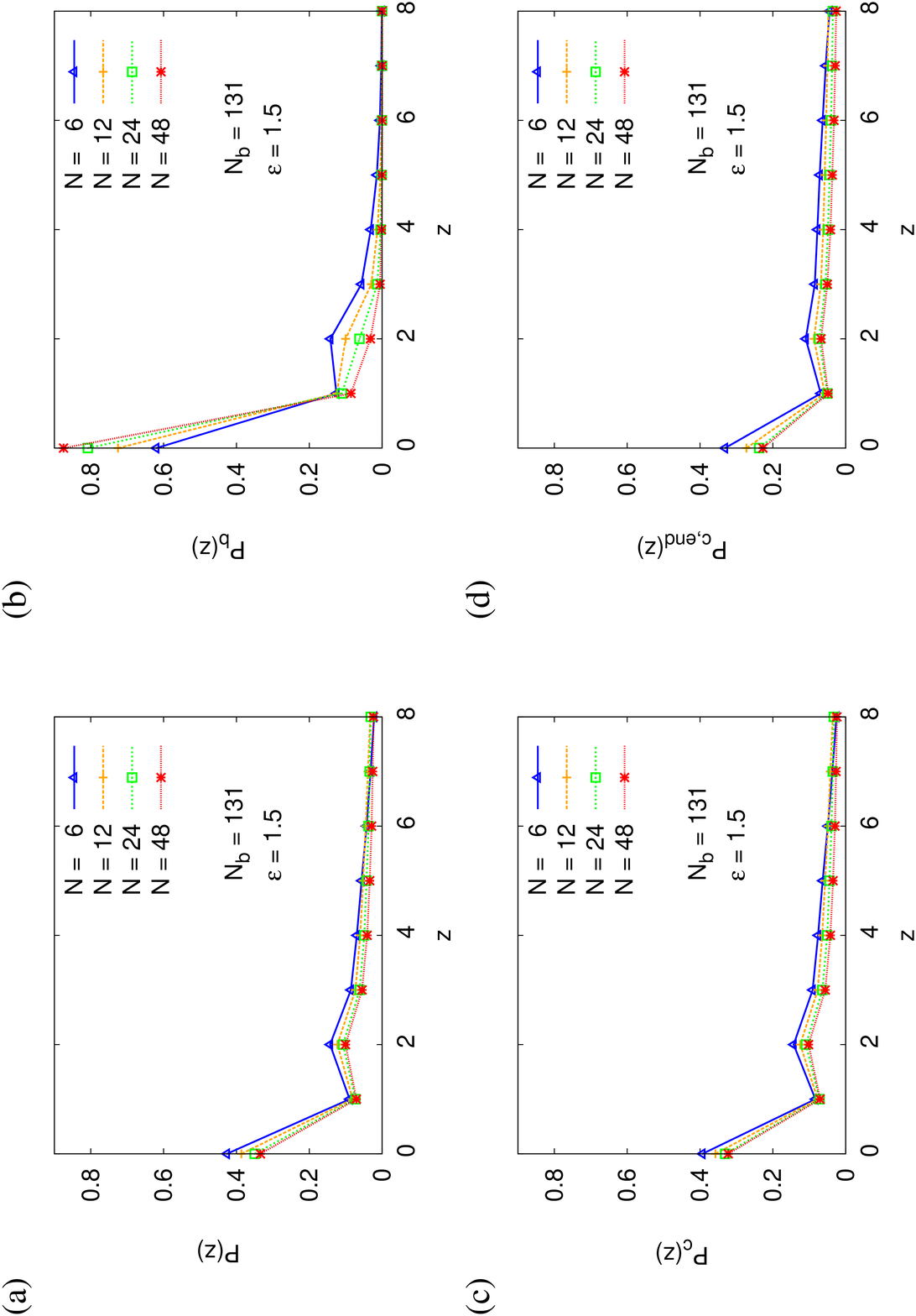}
\caption{\label{fig13}  Same as Fig.~\ref{fig9}, but
for $N_b = 131$ and $\epsilon = 1.50$}
\end{center}
\end{figure*}

\begin{figure*}[htb]
\begin{center}
\includegraphics[scale=0.30,angle=270]{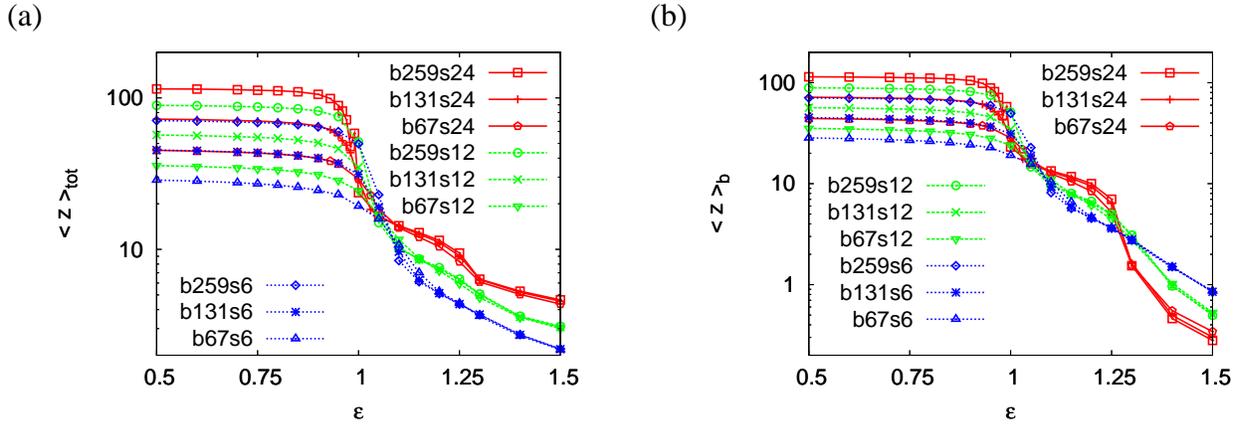}
\caption{\label{fig14} Plot of $\langle z \rangle$
(note the logarithmic scale) vs. $\epsilon$, for various choices
of $N_b$ and $N$ (labeled as $bN_bsN$ in the figure). Case (a)
refers to all monomers of the bottle-brush, case (b) refers to
backbone monomers only.}
\end{center}
\end{figure*}

\begin{figure*}[htb]
\begin{center}
\includegraphics[scale=0.30,angle=270]{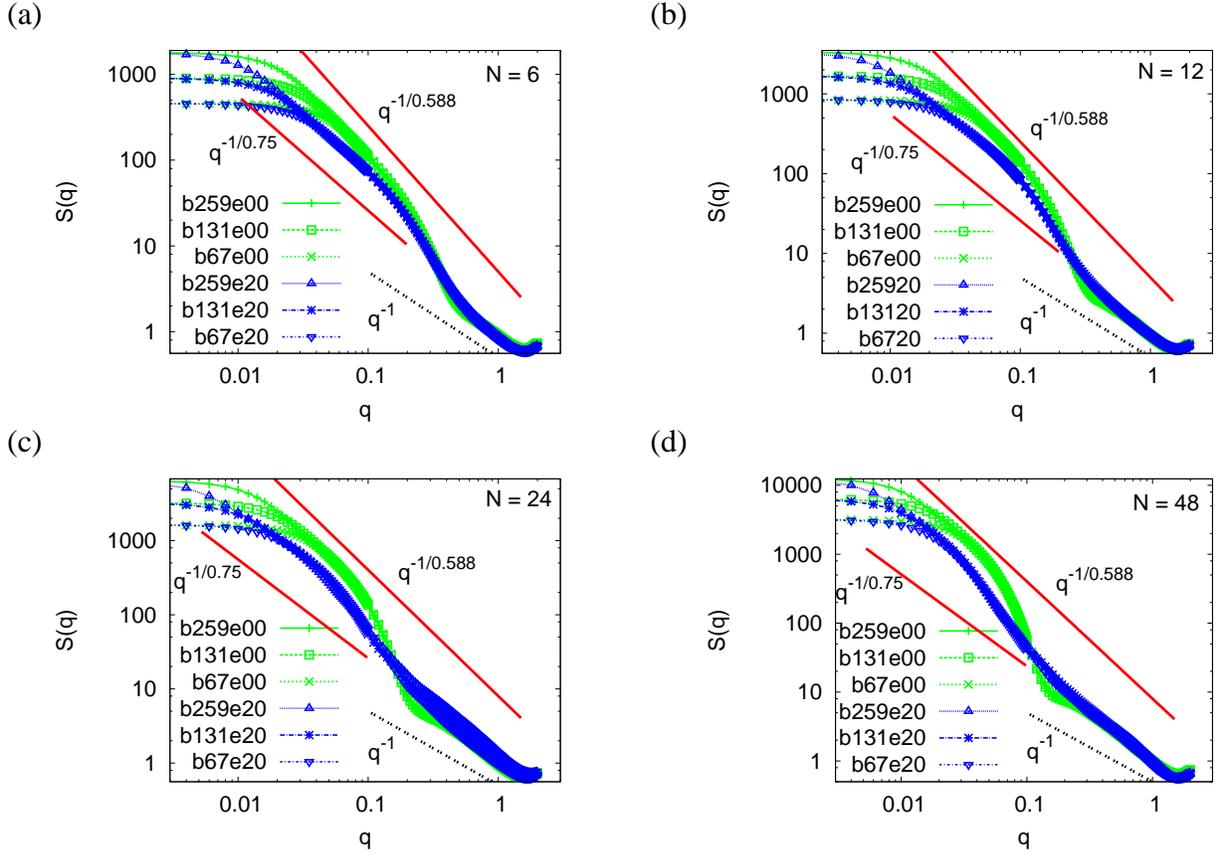}
\caption{\label{fig15} Log-log plot of the scattering
function $S(q)$ of the whole bottle-brush polymers versus wave
number $q$, for side chain lengths $N=6$ (a), $N=12$ (b), $N=24$ (c) 
and $N=48$ (d). Three choices of backbone chain lengths
$N_b=67, 131,$ and 259 are included. Data for $\epsilon = 0.0$ are
denoted as $bN_be00$ while data for $\epsilon = 2.0$ are
denoted as $bN_be20$. Straight lines indicate the power laws
for rods $(q^{-1})$, three-dimensional coils $(q^{-1/\nu}$ with
$\nu = 0.588)$ and two-dimensional coils $(q^{-1/\nu_2},
\nu_2=0.75$), respectively.}
\end{center}
\end{figure*}

\begin{figure*}[htb]
\begin{center}
\includegraphics[scale=0.30,angle=270]{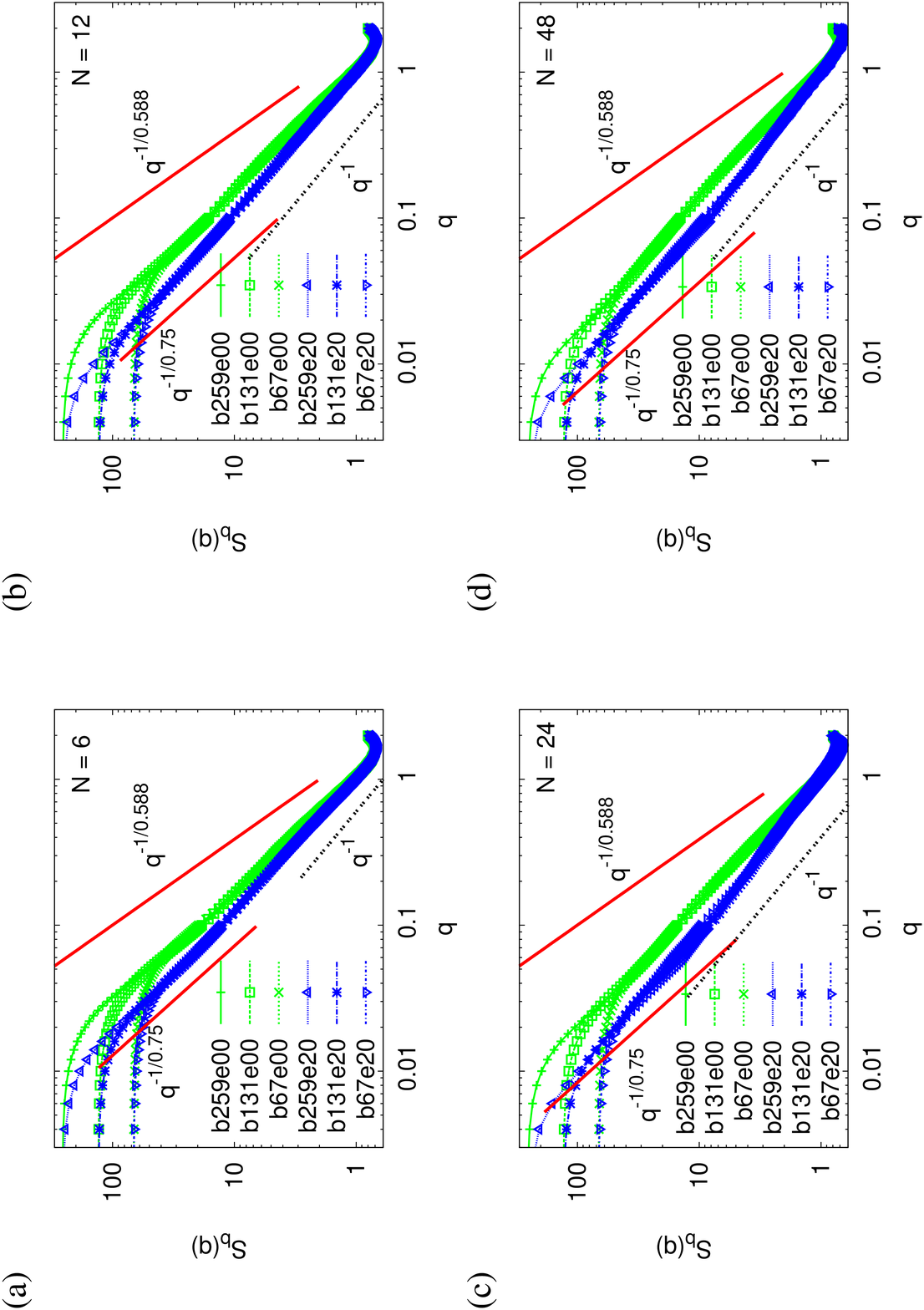}
\caption{\label{fig16} Log-log plot of the scattering
function $S_b(q)$ of the monomers in the  backbone vs. wave number
$q$, for side chain lengths $N=6$ (a), $N=12$ (b), $N=24$ (c) and
$N=48$ (d). Three choices of backbone chain length $N_b=67$, 131,
and 259 are included in the same notation as in Fig.~\ref{fig15} is
used.}
\end{center}
\end{figure*}

A rather unexpected feature, however, emerges when we examine the
variation of the radii with $\epsilon$ (Fig.~\ref{fig8}). 
The side chain linear dimensions are almost independent of $N_b$ for $\epsilon
< 1$. The adsorbtion process for $\epsilon > 1$, however, seems to be an
$N$-dependent two-step process: first a rather rapid decrease of
$R^2_{gc,z}$ (and a rather rapid increase of $R^2_{gc,||}$)
occurs, and the behavior for all three values of $N$ is rather
similar; then, near $\epsilon \approx 1.25$, the behavior changes, and
the further variation depends distinctly on $N$ and for
$R^2_{gc,z}$ and $N=24$ even is non-monotonic. Near $\epsilon
\approx 1.7$ the three sets of curves intersect each other. We
speculate that for $\epsilon < 1.25$ the adsorption is controlled by the side
chains near the grafted end of the 
backbone while for $\epsilon > 1.25$ it is controlled by all
the side chains and the backbone, but clearly this point needs further study. Of
course, $R^2_{gc,z}$ can be considered as a measure of the
thickness of the ``pancake''.

Alternative information on this thickness can be obtained when one
simply studies the monomer density profile $P(z)$, which again can
also be resolved distinguishing contributions from the backbone
only, from the side chains only, or even from the free chain ends
of the side chains only (Figs.~\ref{fig9}-\ref{fig13}).

While for $N=6$ the monomer distribution of the bottle-brush as a
whole for the non-adsorbed bottle-brush mushroom as well as the
distribution of the backbone and the side chain monomers are very
similar to each other and similar to corresponding data for simple
non-adsorbed polymer mushrooms, the distribution gets a more
complicated shape the longer the side chain length $N$ becomes.
The strong shift of the maximum of all distributions with
increasing $N$ to larger $z$ clearly can be attributed to
the increasing backbone stiffness and the corresponding increase
of the linear dimensions of the bottle-brush. When now the
adsorption energy is switched on, the distributions show little
change for $\epsilon \leq 0.5$ (not shown), while near the
adsorption transition a profound change of the character of the
distribution has occurred (Fig.~\ref{fig11}): all distributions
now exhibit a second maximum near the surface, for large enough
$N$, and also the main maximum occurs now much closer to the
adsorbing surface (but its position $z_{\textrm{max}}$ strongly
increases with $N$, presumably because the bottle-brush needs to
avoid ``crowding'' of monomers near the adsorbing surface). In the
adsorbed regime (Figs.~\ref{fig12}, \ref{fig13}) one observes
layering type oscillations near the wall, similar to the behavior
of off-lattice models for fluids close to hard walls. While the
backbone monomers still have a pronounced maximum in $P(z)$ away
from the surface, Fig.~\ref{fig12}(b) the corresponding results
for the side-chain monomers do not show this maximum any more. This
observation corroborates our interpretation of the non-monotonic
variation of $R^2_{gc,z}$ in Fig.~\ref{fig8}, namely that
adsorption occurs in two steps - first the adsorption of the
monomers near the grafted end of the backbone takes place, while the adsorption
transition of the rest of the bottle-brush is not
yet completed at $\epsilon = 1.25$ for the side chain lengths $N$
studied here. For $\epsilon = 1.5$, on the other hand, all
profiles show only very little dependence on the side chain length
(Fig.~\ref{fig13}): The lengths of still non-adsorbed ``tails''
and ``loops'' \cite{27} then are rather short, and therefore the
side chain length no longer is important.

From the profiles, one can easily estimate the average thickness
$\langle z \rangle$ defined as $\langle z \rangle = \int \limits
_0 ^\infty zP(z) dz / \int \limits _0^\infty P(z)dz$, and study
its dependence on $\epsilon$ (Fig.~\ref{fig14}).
Again the two-step character of adsorption for long enough
side-chains $(N=24)$ is very clearly seen.

As a final point of this study, we address the scattering function
$S(q)$ where the orientation of the wave vector $\vec{q}$ has been
averaged over. Hence
\begin{equation}\label{eq5}
S(q) = \frac {1}{\mathcal{N}^2} \langle \sum \limits
_{j=0}^{\mathcal{N}} \sum \limits _{k=0}^{\mathcal{N}} \exp [i
\vec{q}\cdot (\vec{r}_j - \vec{r}_k)] \rangle
\end{equation}
where $\mathcal{N}$ is the total number of monomers, from which
the scattering is considered, and the average $\langle
\ldots\rangle$ includes both a statistical average over the
conformations of the bottle-brush and a spherical average over the
direction of $\vec{q}$. A distinctive advantage of the simulation
is that the scattering from the total bottle-brush, the scattering
only from the backbone, or only from the side chains are easily
accessible.

Fig.~\ref{fig15} correspondingly compares data for $S(q)$
considering the total scattering from all the monomers of the
bottle-brush, comparing data for 4 choices of $N$ and 3 choices of
$N_b$, as indicated, and including data for both $\epsilon = 0.0$
and $\epsilon = 2.0$. While the data for $\epsilon = 0.0$ closely
resemble the scattering function of isolated bottle-brush polymers
in dilute bulk solution, as studied previously, revealing a law
$q^{-1/\nu}$ for small $N$ and intermediate $q$, and a rod-like
behavior $q^{-1}$ for large $q$, for the strongly adsorbed case 
$(\epsilon = 2.0)$ for large $q$ the rod-like behavior is not seen for $N=24$
and $N=48$, rather one finds in the decade $0.1 \leq q \leq 1.0$ a
behavior proportional to $q^{-1/\nu_2} = q^{-4/3}$: the side
chains in the strongly adsorbed case behave like two-dimensional
self-avoiding walks: since there so many more monomers occur in
the side chains than in the backbone, the rod-like characteristics
of the latter is only revealed when one focuses on the scattering
from the backbone only (Fig.~\ref{fig16}). Then one can clearly
see two crossovers, $S_b(q) \approx N_b(1-q^2 \langle
R_{g,b}^2 \rangle /3)$ at small $q$ of order $\langle R^2_{g,b} \rangle ^{-1}$ 
crosses over to $q^{-4/3}$ in $d= 2$ and to
$q^{-1/0.588}$ in $d=3$ dimensions, while near $q \approx 0.1$ a
crossover to $S_b(q) \propto q^{-1}$ occurs, both for adsorbed and
for non-adsorbed bottle-brushes.

\section{SUMMARY}

In this paper a Monte Carlo simulation study of the adsorption of
bottle-brush mushrooms'' (i.e. bottle-brush polymers with a
backbone chain end grafted to a flat structureless impenetrable
surface) has been presented, using the bond-fluctuation model and
assuming very good solvent conditions. The same range of backbone
chain lengths $(N_b \leq 259)$ and side chain lengths $(N \leq
48)$ as used in a previous study of the same model in dilute
solution in the bulk \cite{19} has been used, since evidence has
been presented that this range is fully appropriate to allow a
comparison with experiment \cite{19}. Every backbone monomer
carries one side chain.

Both the backbone chain and the side chains are assumed to be
fully flexible, and a short range attractive energy $\epsilon$
(putting temperature $k_BT=1$ throughout) is assumed that attracts
both monomers of the side chains and of the backbone to the
surface in the same way. We show that near $\epsilon_c(N)=1$ the
mushrooms cross over from a three-dimensional configuration (for
$\epsilon <1$) to the adsorbed, quasi-two-dimensional configuration for
$\epsilon >1$. These transitions occur roughly at the same value
of $\epsilon _c(N)$ irrespective of $N$, but the latter has a
pronounced effect deeper in the adsorbed region, where chain
linear dimensions reveal a two-step adsorption process
(Figs.~\ref{fig8}, \ref{fig14}). 
We tentatively associate the first step of the
adsorption (near $\epsilon_c$) to the side chains near the grafted end of the
backbone, while the backbone as a whole for 
$\epsilon$ not much larger than $\epsilon_c$ exhibits only a very
small fraction $N_{s,b}/N_b$ of adsorbed monomers
(Fig.~\ref{fig3}(c)). Thus, most of the backbone monomers still
occur in the tail or in rather long loops stretching away from the adsorbing
surface. Only at somewhat larger values of $\epsilon$ (namely for
$\epsilon \approx 1.2$ in our case) the ``trains'' \cite{27} of
consecutively adsorbed backbone monomers get longer and more
frequent, and the tail and the loops get correspondingly shorter. This behavior
must cause an interesting interplay with the behavior of the side
chains: a side chain grafted to a backbone monomer that is part of
an adsorbed ``train'' behaves basically like a small mushroom, and
hence such a side chain gets easily adsorbed. However, side chains
grafted to a backbone monomer that belongs to the tail or to a large loop cannot
yet get adsorbed easily, simply because typically the grafting
site of this side chain is not close enough to the adsorbing
surface. However, when $\epsilon$ increases and the loops in the
backbone would get smaller, the grafting site gets close to the
adsorbing surface, and then the side chain easily can adsorb (and
if it is long enough, the side chain adsorption will have a kind
of feedback effect on the backbone, dragging the remaining part of
the non-adsorbed loop towards the surface as well). Clearly, this
picture is qualitative and somewhat speculative, but it suggests
that through a cooperative interplay of side chains and backbone a
much more intricate behavior is possible than for the adsorption
of linear polymers. 
One can even speculate that due to the strong crowding effects upon adsorption,
occurring especially for long side chains, a complete adsorption of the
bottle-brush can not occur from a configuration where both ends are
already adsorbed and a train is still present in the interior of the
polymer. Rather the adsorbed part towards the free end of the backbone would
have do desorb again first and then the complete adsorption proceeds in a
zipper-like fashion starting from the grafted chain end.
Of course, development of an analytical model
that could describe data such as shown in Figs.~\ref{fig2},
\ref{fig3}, \ref{fig8} and \ref{fig14} would be highly desirable.

We have also verified that strongly adsorbed bottle-brush
polymers exhibit lateral linear dimensions that scale with the
exponent $\nu_2$ of two-dimensional self-avoiding walks, and in
the scattering function of the total bottle-brush this behavior is
seen as well (Fig.~\ref{fig15}). However, the expected rod-like
behavior on intermediate length scales is only seen when the
scattering from the backbone is isolated (Fig.~\ref{fig16}). Our
data also imply rather different structures of adsorbed vs.
non-adsorbed bottle-brush polymers making it difficult to reason from adsorbed
structures on the behavior of free
bottle-brushes in solution! We emphasize this simple point, as a
warning to premature interpretations of corresponding experiments.

Our study reveals many details (bimodal behavior of the density
distribution $P(z)$ of the monomers as a function of distance from
the surface near adsorption (Fig.~\ref{fig12}), characteristic
effects due to the backbone chain ends (Figs.~\ref{fig4} -
\ref{fig7}), etc., which all reflect in some way the interplay
between the enthalpy won by adsorbing monomers, and various
entropic effects. More work is clearly required to clarify the
reasons for these detailed observations. Also, it would be very
interesting to clarify the nature of the cross-sectional structure
of the adsorbed bottle-brush on a coarse-grained scale (in the
bulk, the bottle-brush can be viewed as a more or less flexible
cylinder: should we view the cross section of an adsorbed bottle
brush as a sphere cap rather than a sphere?) A further interesting
aspect is the question of the persistence length of adsorbed
bottle-brushes \cite{potemkin,feuz}, and its relation to the persistence length of
bottle-brushes in the bulk. 
Other interesting issues concern the effect of solvent quality~\cite{42}
and possible irreversibility effects in the adsorption of 
the side arms~\cite{43}.
Our study only considers adsorption under very good solvent
conditions. It also is a very interesting problem
(and relevant for experiment) to consider adsorption under poor
solvent conditions where much more compact ``pancake" structures
should result~\cite{42} rather than the configurations obtained here,
that resemble two dimensional self-avoiding walks.
Finally, we mention the possibility that for strong adsorption
$(\epsilon \gg \epsilon_c)$ the left-right distribution of the 
side chains with respect to the backbone is quenched, i.e., random
fluctuations in this distribution have not enough time to relax.
Such effects are predicted to have interesting effects on the conformation
of such strongly adsorbed bottle-brush polymers~\cite{43}.
We intend to study some of these
issues in forthcoming work.

\begin{acknowledgments} H.-P.~H. received funding from the
Deutsche Forschungsgemeinschaft (DFG), grant No SFB 625/A3. We are
grateful for extensive grants of computer time at the JUROPA 
under the project No HMZ03 and
SOFTCOMP computers at the J\"ulich Supercomputing Centre (JSC).
\end{acknowledgments}

\appendix
\section{Scaling Analysis of the adsorption transition bottle-brush polymers}

In the previous work where the adsorption transition of flexible
chains end-grafted by a chain end on an impenetrable surface
was studied by Monte Carlo methods~\cite{26,29,31,31a},
the location of the transition in the limit of infinite number of
bonds between the monomers, $N_b \rightarrow \infty$, typically
has been extracted from a scaling analysis~\cite{26}. The
statement of scaling is that quantities such as the fraction
$N_s/N_b$ of adsorbed monomers do not depend on the two
variables $N_b$, $\kappa=(\epsilon-\epsilon_c)/\epsilon_c$
in the most general way, but are essentially a function of
a simple scaling variable, $\zeta=\kappa N_b^{\phi}$,
apart from a power law prefactor. Thus
\begin{equation}
    N_s = N_b^\phi F_s(\zeta)\;, \enspace \kappa \rightarrow 0\;,
\, N_b\rightarrow \infty \;.
\label{eqa1}
\end{equation}
The power laws quoted in the main text \{Eq.~(\ref{eq1})\} simply
result from Eq.~(\ref{eqa1}) for $\zeta \rightarrow -\infty$,
$\zeta =0$, and $\zeta \rightarrow +\infty$. Similar scaling
results hold for the linear dimensions of the chains.
E. g. the mean square gyration radius components of the
backbone of the bottle-brush perpendicular and parallel to
the surface are
\begin{equation}
    R_{gb,z}^2 = N_b^{2 \nu} F_{bz}(\zeta) \;, \,
R^2_{gb,||}=N_b^{2 \nu} F_{b ||}(\zeta) \;,
\label{eqa2}
\end{equation}
and again Eq.~(\ref{eqa2}) is supposed to be valid in the limit
where both $\kappa \rightarrow 0$ and $N_b \rightarrow \infty$.
For Eq.~(\ref{eqa2}), the scaling functions for $\zeta \le 0$
are less interesting, since $F_{bz}(\zeta \rightarrow -\infty)$,
$F_{b||}(\zeta \rightarrow -\infty)$, $F_{bz}(0)$, and $F_{b||}(0)$
are finite constants \{cf. Eq.~(\ref{eq2})\}. However, for
$\zeta \rightarrow +\infty$ one has
\begin{equation}
  F_{bz}(\zeta) \propto \zeta^{-2 \nu/\phi} \;, \,
F_{b||}(\zeta) \propto \zeta^{2(\nu_2-\nu)/\phi} \;,
\label{eqa3}
\end{equation}
and combining Eqs.~(\ref{eqa2}), (\ref{eqa3}) one hence recovers
Eqs.~(\ref{eq3}), (\ref{eq4}) of the main text.

\begin{figure*}[htb]
\begin{center}
\includegraphics[scale=0.30,angle=270]{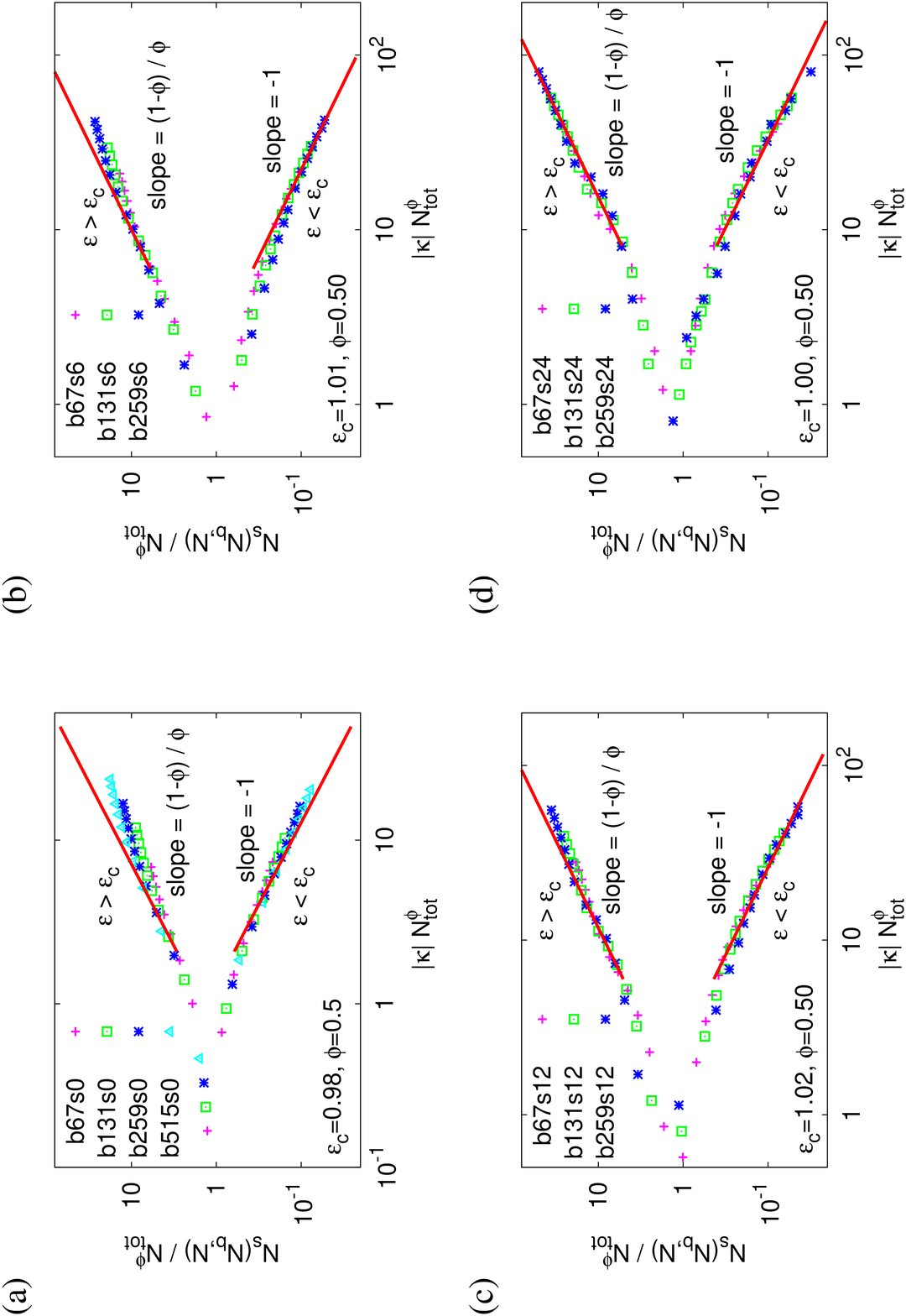}
\caption{\label{figa1} 
Log-log plot of the rescaled surface contacts $N_s/N_{\rm tot}^{\phi}$
vs. $\mid \kappa \mid N_{\rm tot}^{\phi}$ for $\phi=0.50$.
Data for backbone lengths
$N_b=67$, $131$, $259$, and $515$ are included, as well as side
chain lengths (a) $N=0$, (b) $N=6$, (c) $N=12$, and (d) $N=24$.}
\end{center}
\end{figure*}

\begin{figure*}[htb]
\begin{center}
\includegraphics[scale=0.30,angle=270]{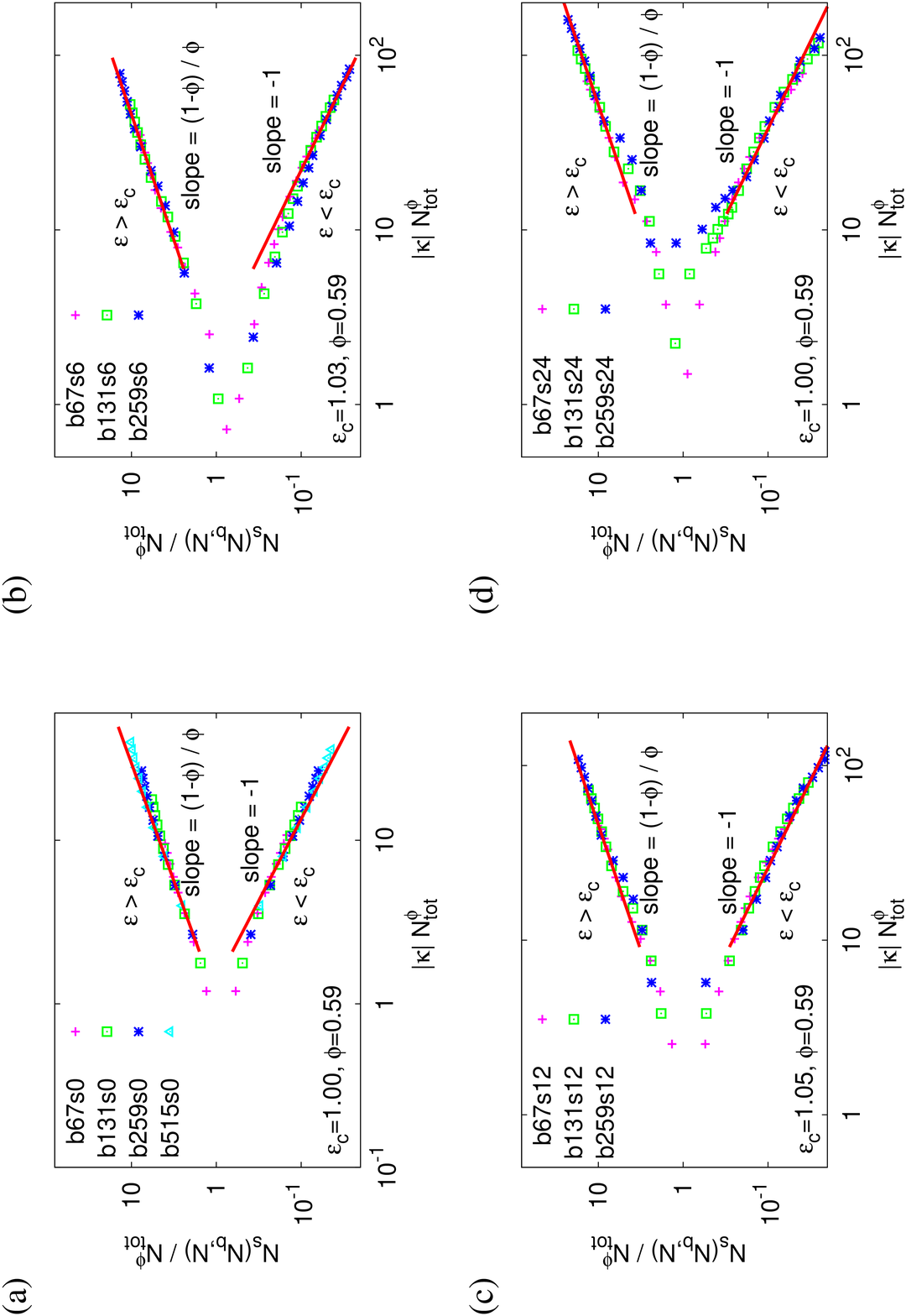} 
\caption{\label{figa2} 
Log-log plot of the rescaled surface contacts $N_s/N_{\rm tot}^{\phi}$
vs. $\mid \kappa \mid N_{\rm tot}^{\phi}$ for $\phi=0.59$.
Data for backbone lengths
$N_b=67$, $131$, $259$, and $515$ are included, as well as side
chain lengths (a) $N=0$, (b) $N=6$, (c) $N=12$, and (d) $N=24$.}
\end{center}
\end{figure*}

\begin{figure*}[htb]
\begin{center}
\includegraphics[scale=0.30,angle=270]{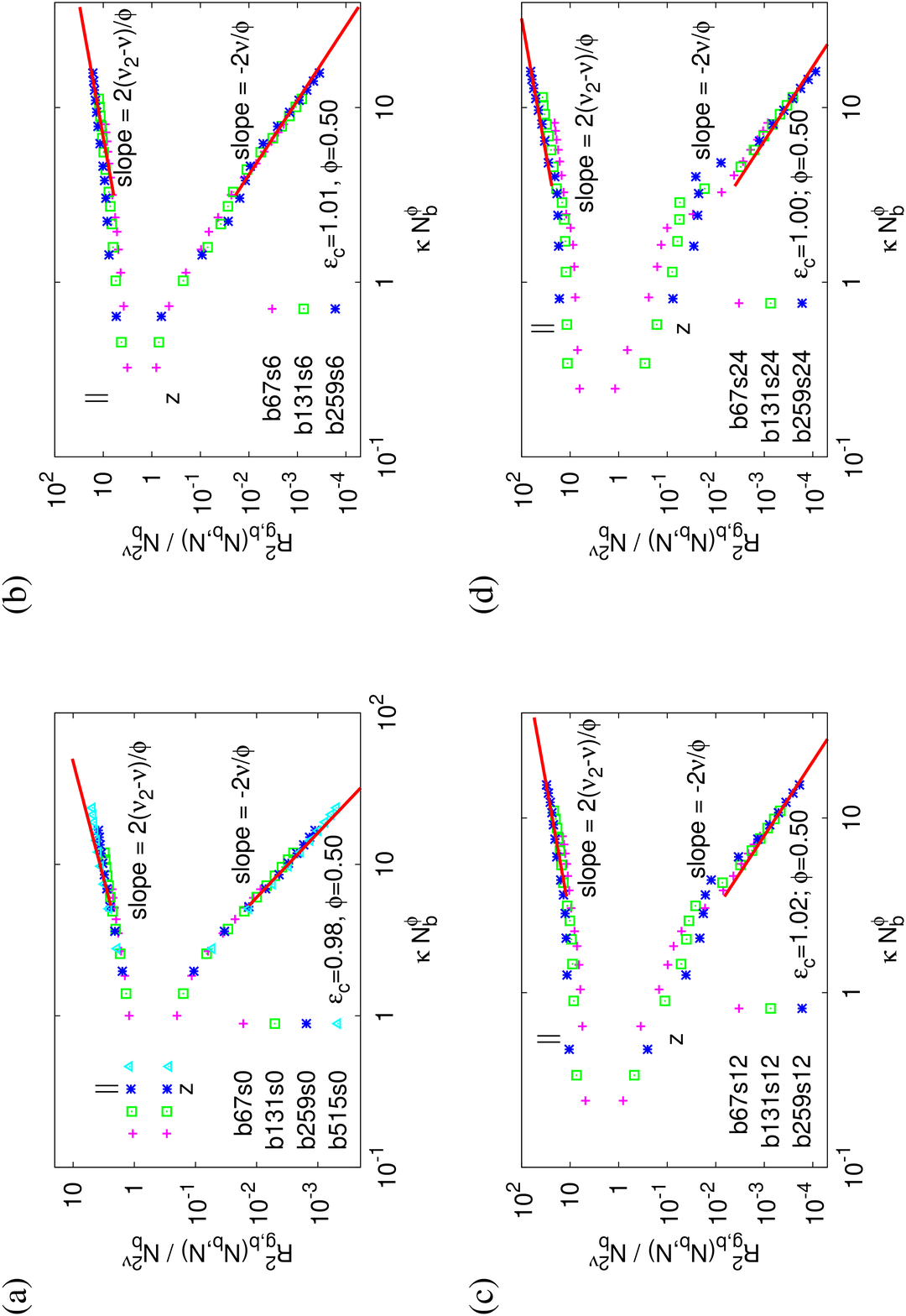}
\caption{\label{figa3} 
Log-log plot of the mean square gyration radius components
perpendicular and parallel to the surface, $R^2_{gb,z}/N_b^{2\nu}$
and $R^2_{gb,||}/N_b^{2\nu}$, respectively,
vs. $\kappa N_{\rm tot}^{\phi}$ for $\phi=0.50$.
Data for backbone lengths
$N_b=67$, $131$, $259$, and $515$ are included, as well as side
chain lengths (a) $N=0$, (b) $N=6$, (c) $N=12$, and (d) $N=24$.
Here the Flory exponents $\nu_2=3/4$ (2D), and $\nu=0.588$ (3D).}
\end{center}
\end{figure*}

\begin{figure*}[htb]
\begin{center}
\includegraphics[scale=0.30,angle=270]{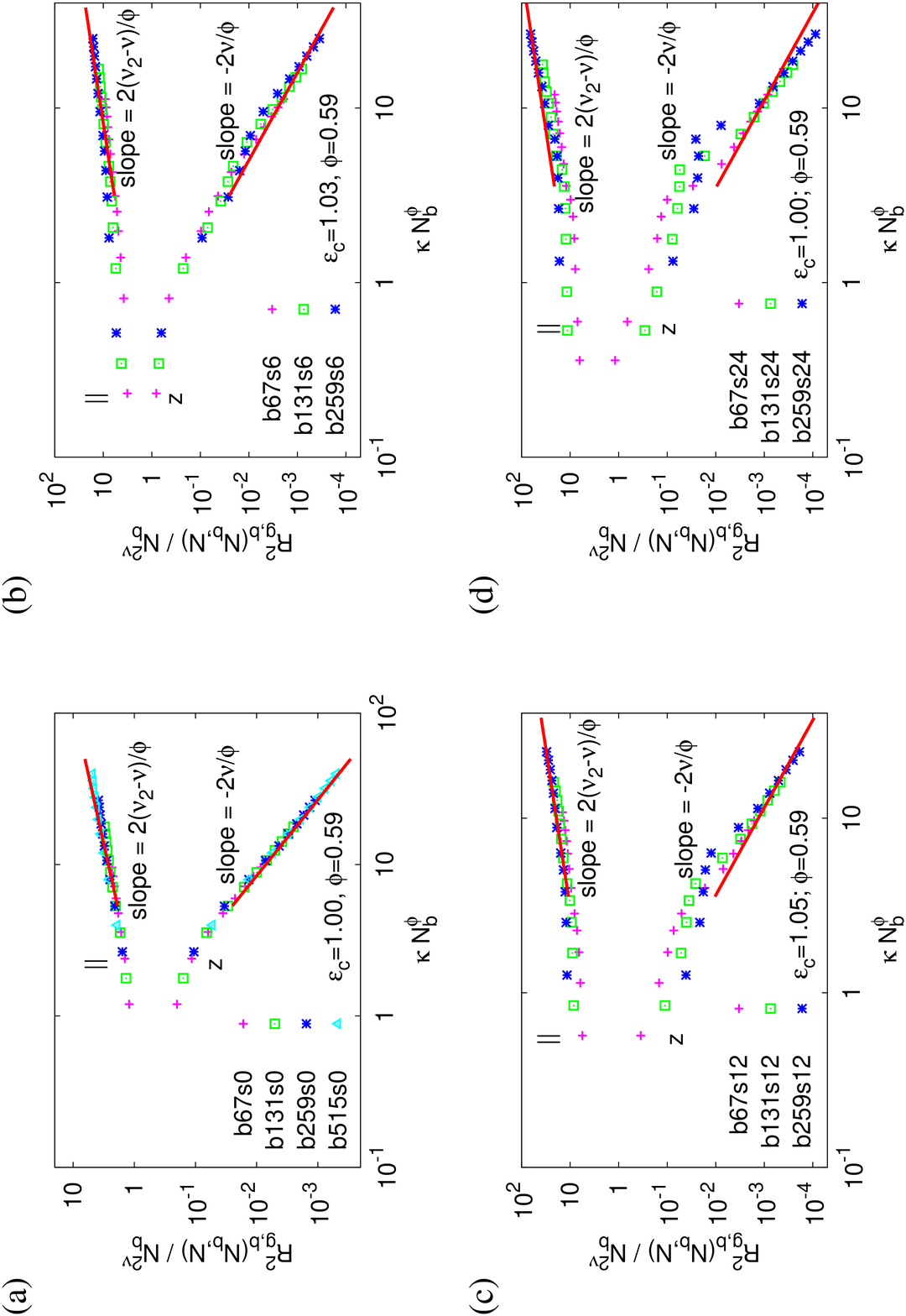} 
\caption{\label{figa4} 
Log-log plot of the mean square gyration radius components
perpendicular and parallel to the surface, $R^2_{gb,z}/N_b^{2\nu}$
and $R^2_{gb,||}/N_b^{2\nu}$, respectively,
vs. $\kappa N_{\rm tot}^{\phi}$ for $\phi=0.59$.
Data for backbone lengths
$N_b=67$, $131$, $259$, and $515$ are included, as well as side
chain lengths (a) $N=0$, (b) $N=6$, (c) $N=12$, and (d) $N=24$.
Here the Flory exponents $\nu_2=3/4$ (2D), and $\nu=0.588$ (3D).}
\end{center}
\end{figure*}

Of course, the singularities described by Eqs.~(\ref{eq1})-(\ref{eq4})
of the main text and the above Eqs.~(\ref{eqa1})-(\ref{eqa3})
emerge only in the double limit $\kappa \rightarrow 0$,
$N_b \rightarrow \infty$: for any finite $N_b$ functions
$N_s(\epsilon)$, $R^2_{gb,z}(\epsilon)$ and $R_{gb,||}^2(\epsilon)$
as regular functions of $\epsilon$, the singular behavior is rounded
off, as is well known~\cite{26}. Thus the estimation of $\epsilon_c$
is a nontrivial matter. What usually is done, see 
e.g. Refs.~\cite{26,31,31a}, is a ``data collapsing method":
one varies both $\epsilon_c$ and the (originally unknown)
value of the crossover exponent to obtain an optimal
``data collapse" of the set of functions $N_s(\epsilon, N_b)$,
$R^2_{gb,z}(\epsilon, N_b)$ and $R^2_{gb,||}(\epsilon,N_b)$ on these 
``master curves" which represent then the three scaling functions
$F_s(\zeta)$, $F_{bz}(\zeta)$, and $F_{b||}(\zeta)$.
Figs.~\ref{figa1}-\ref{figa4} show some typical attempts to do 
this with our data (each figure has four parts to show the results
for the four side chain lengths $N=0$ (no side chains),
$N=6$, $N=12$, and $N=24$, respectively). Here we have used as 
a further constraint the ``universality principle", namely
the crossover exponent $\phi$ should be a universal constant,
independent of irrelevant microscopic details such as the 
length $N$ of the side chains (as long as $N$ is finite,
while the limit $N_b \rightarrow \infty$ is considered).
Note also that the ratio $R^2_{gb,z}/R^2_{gb, ||}$ according to 
Eq.~(\ref{eqa2}) should be a function of $\zeta$ alone, and plotted
vs. $\epsilon$ should have a unique intersection point at
$\epsilon_c$. (However, as shown in Fig.~\ref{fig034}, 
one would need data for larger $N_b$ and very good statistical
accuracy to show this).

   The straight lines included in these log-log plots
of the resulting estimates for the scaling functions that they
must converge to for $\mid \zeta \mid \rightarrow \infty$
yield the power laws quoted in Eqs.~(\ref{eq1})-(\ref{eq4}).
Since $\nu=0.588$ and $\nu_2=3/4$ is independently
known, it is again only the exponent $\phi$ (as in the scaling
variable $\kappa N_b^\phi$) which matters. To simplify matters,
we only present the two extreme choices for $\phi$ here:
the estimate $\phi=0.5$ (which happens to coincide with
the crossover exponent for adsorption of Gaussian chains~\cite{26,27,28},
but is supported by a renormalization group estimation~\cite{30}
and some Monte Carlo studies~\cite{29}) and the most recent estimate
$\phi=0.59$ from Monte Carlo studies~\cite{31a}. Comparison of 
Figs~\ref{figa1}, \ref{figa2} for $N_s$ shows that we confirm
the finding of Descas et. al.~\cite{31a}, that
$\phi=0.59$ yields a slightly better data collapse on the 
master curve $F_s(\zeta)$, $\epsilon_c(N)$ being always
chosen such that for $\epsilon < \epsilon_c$ the master curve
has the correct slope. For $\epsilon > \epsilon_c$ the data
for large $\zeta$ fall systematically somewhat below the expected
power law, if $\phi=0.5$ is chosen. However, as a caveat
we mention that for finite $N_b$ the data are not expected  
to follow the master curve for very large $\zeta$, since 
$N_s/N_b \rightarrow 1$ (for the case $N=0$ and large enough 
$\epsilon$, where the chain is adsorbed in a two dimensional
configuration). So, it must happen that for large $\zeta$
the curves bend away from the power law that describe
the asymptotic limit of the scaling function,
and this ``saturation effect" occurs the later the larger $N_b$
is, and this is exactly what one sees in Fig.~\ref{figa1}(a), 
while in Fig.~\ref{figa2} (a) the same effect is somewhat less
pronounced. However, in our opinion the evidence from 
these plots that $\phi=0.59$ is ``better" than $\phi=0.50$ is
somewhat weak, and much longer chains would be needed to reach
a really firm conclusions.

  When one chooses $\phi=0.50$, the resulting estimates for 
$\epsilon_c(N)$ would be $0.98$, $1.01$, $1.02$, and $1.00$, 
for side chain lengths $N=0$, 6, 12, 24, respectively.
When one chooses $\phi=0.59$, on the other hand, most of the
estimate are a little bit larger, namely 1.00, 1.03, 1.05,
and 1.00, respectively. As expected, the best estimates for $\phi$
and $\epsilon_c(N)$ are correlated. However, it must be
admitted that slightly different choices of $\epsilon_c(N)$
than those that are shown here yield a data collapse that is only slightly
worse. Given the fact that Eqs.~(\ref{eq1})-(\ref{eq4}) and
(\ref{eqa1})-(\ref{eqa3}) are only asymptotically valid in the 
double limit $\kappa \rightarrow 0$ and $N_b \rightarrow \infty$,
while for finite $N_b$ and finite nonzero $\kappa$ corrections to
scaling may be present, a reliable judgment of accuracy for
the estimates $\epsilon_c(N)$ is difficult. For this reason,
we have quoted $\epsilon_c \approx 1.0$ as an estimate for all
choices of $N$ studied in the main text.

  In fact, the estimation is not really improved when one considers
the scaling of the mean square gyration radius 
(Figs.~\ref{figa3}, \ref{figa4}).
For $N=0$ and $N=6$, the quality of the data collapse for 
$\phi=0.50$ and $\phi=0.59$ is of comparable quality. For $N=12$
and $N=24$, scaling seems to work well only for large $\zeta$,
not for small $\zeta$, indicating clearly a more complex
behavior right at the adsorption transition. The data show that
the choices for $\epsilon_c(N)$ quoted above are compatible with 
the behavior of the chain dimensions too, at least roughly.

  The result that the addition of side chains (that are still short,
of course) has so little effect on the location of the adsorption
transition clearly is surprising and unexpected. Therefore we have
presented in this Appendix our evidence for this fact in some detail, 
not withstanding the uncertainty about the best value of 
the crossover exponent; this problem is not solved by
our analysis.


\begin{thebibliography}{99}
\bibitem{23} Note that previous work on the adsorption transition
of comb polymers used $N = 5$ only \cite{21}.
\end{thebibliography}
\end{document}